# Epitaxial ferroelectric $HfO_2$ films: growth, properties and devices

Ignasi Fina[a] and Florencio Sánchez[b]

Institut de Ciència de Materials de Barcelona (ICMAB-CSIC), Campus UAB, Bellaterra 08193, Spain

[a] ifina@icmab.es, [b] fsanchez@icmab.es

**Abstract**
About ten years after ferroelectricity was first reported in doped $HfO_2$ polycrystalline films, there is tremendous interest in this material and ferroelectric oxides are once again in the spotlight of the memories industry. Great efforts are being made to understand and control ferroelectric properties. Epitaxial films, which have fewer defects and a more controlled microstructure than polycrystalline films, can be very useful for this purpose. Epitaxial films of ferroelectric $HfO_2$ have been much less investigated, but after the first report in 2015 significant progress has been achieved. This review summarizes and discusses the main advances on epitaxial $HfO_2$, considering growth, study of structural and ferroelectric properties, identification of the ferroelectric phase, and fabrication of devices. We hope this review will help to researchers investigating epitaxial $HfO_2$. It can also help extend the interest of the ferroelectric $HfO_2$ community, now basically focused on polycrystalline samples, to epitaxial films.

## 1. Introduction

Ferroelectric perovskites are widely used in the electronics industry, in particular $BaTiO_3$ is used in multilayer ceramic capacitors and lead zirconate titanate (PZT) in piezoelectric devices.[1] Additionally, ferroelectric random access memories (Fe-RAM), which currently use PZT, are on the market for about 30 years. These memories offer very high speed, low power consumption, and high retention and endurance. However, the difficulties in scaling the size below the 130 nm node make them not cost competitive and its use is restricted to niche applications. Various perovskite oxides are also used in emerging memories such as ferroelectric field effect transistors (Fe-FETs) and ferroelectric tunnel junctions (Fe-TJs),[2] but after several decades of research, they are far from being used in commercial devices. Therefore, ferroelectric perovskites do not have a big impact on the memory device market. Limited success stems from the complexity of these oxides. Ferroelectric perovskite films must be no less than tens of nanometer thick to exhibit good properties, and deposition techniques as atomic layer deposition (ALD) allowing conformal growth are not easily implemented. The compatibility with CMOS processes is restricted by the thermal budget required for its crystalline growth and by its sensibility to hydrogen.

The recent discovery of ferroelectricity in simple oxides based on $HfO_2$ has renewed the interest in ferroelectric memories.[2-7] $HfO_2$ is already used in CMOS lines and the deposition techniques and conditions to form the ferroelectric phase can be compatible with CMOS technology. Very shortly after the first publication in 2011 reporting on ferroelectricity in $HfO_2$,[8] prototypes of Fe-RAMs and Fe-FETs[9] were produced. This was followed by the fabrication of 3D Fe-RAMs[10] and Fe-FETs with 22 nm node.[11] Doped-$HfO_2$ is also promising for the development of Fe-TJs.[12] Research on $HfO_2$ has basically been done with polycrystalline films, mostly made by ALD at room temperature followed of rapid thermal annealing using TiN/doped-$HfO_2$/TiN capacitor heterostructures.

The experimental demonstration of ferroelectricity in $HfO_2$ was unexpected since its monoclinic phase (space group $P2_1/c$) is nonpolar. Other crystalline phases that form at high temperature, the tetragonal ($P4_2/nmc$) and cubic (Fm3m) polymorphs, are also nonpolar. It was observed that the presence of a capping layer before $HfO_2$ annealing was essential to stabilize the ferroelectric phase. It was proposed that the film crystallized in tetragonal phase, transforming into an orthorhombic polar phase during cooling. In absence of the capping layer, it transformed into the monoclinic phase.[8, 13] The films with capping layer and annealed were not pure orthorhombic and the monoclinic phase was also present. The relative amount of the monoclinic phase increases with thickness in films thicker than around 10 nm.[14-16] The experimental results suggested that surface and strain energy were important factors in the stabilization of the metastable ferroelectric phase. Experimental structural characterization indicated that the $Pca2_1$ phase was the origin of the ferroelectricity in $HfO_2$,[17] and density functional theory (DFT) calculations support this.[18-22] However, the formation of the ferroelectric phase does not depend only on thermodynamics, and the kinetics in the preparation process seems to be determinant.[23]

Progress in the preparation, understanding of structural and ferroelectric properties of $HfO_2$, and fabrication of devices has been enormous (see, for instance, reviews in refs. [3-7, 24-31]). This progress has been achieved mainly with polycrystalline films. Epitaxial films have been much less investigated. The polycrystalline ferroelectric films are formed by annealing of amorphous or quasi-amorphous doped $HfO_2$ in heterostructures such as TiN/doped-$HfO_2$/TiN. Instead, epitaxial films generally nucleate on the surface of single crystals or bottom epitaxial layers. The relative contribution of the interface energy during nucleation is greater than in the case of crystallization in bulk by annealing. This has allowed the epitaxial stabilization of metastable phases of other materials that rarely form in polycrystalline films.[32] Epitaxial films can also undergo phase transformations when cooled, but the contribution of the (semi)coherent interfaces can be much relevant than in polycrystalline films. Therefore, the stabilization mechanisms of the ferroelectric phase of $HfO_2$ in epitaxial films may differ significantly from those widely investigated for polycrystalline films, offering the opportunity to stabilize films of enhanced quality and concomitantly better functional properties. Epitaxial films, with better control of microstructure, can also be of great interest to better understand the ferroelectric properties of $HfO_2$, as well as for prototyping devices. Despite this interest, it was not until 2015 when researchers at Tokyo Institute of Technology and Tohoku University reported by the first time on epitaxial growth of ferroelectric films of doped $HfO_2$.[33] Afterwards the research on epitaxial ferroelectric $HfO_2$ films increased exponentially. Table I summarizes the ferroelectric properties reported for epitaxial films of doped $HfO_2$. Excellent properties have been achieved, and differences between polycrystalline and epitaxial films have been observed (Table II). Here, we review on the main results obtained in epitaxial ferroelectric $HfO_2$ films. We will first revise results on the epitaxial films grown on yttria-stabilized zirconia, epitaxial films obtained by the crystallization of amorphous films, epitaxial films grown on perovskite substrates and, finally, epitaxial films grown on Si. We will discuss on the different polymorphs that have been proposed as ferroelectric phase in epitaxial films. We will analyze in detail the functional properties reported for epitaxial films, i.e. the $P_r$ and $E_c$ obtained values and the endurance and retention. Finally, we will focus on the results obtained in the characterization of devices made of epitaxial doped $HfO_2$ films.

## 2. Epitaxial ferroelectric doped $HfO_2$ films on yttria-stabilized zirconia

Shimizu et al.[33] prepared Y-doped (0, 7, and 15 %) $HfO_2$ (YHO) films, 20 nm thick, on (001)-oriented yttria-stabilized zirconia (YSZ) by pulsed laser deposition (PLD) at 700 °C and 0.01 Torr of $O_2$. Symmetric X-ray diffraction (XRD) measurements signaled that the films presented (001) orientation, as the substrate. The close position of the XRD symmetric reflections made phase identification difficult. The authors used a two-dimensional (2D) detector to measure asymmetric reflections. There are additional weak diffraction spots as lower the crystal symmetry is, and the absence of some reflections permits excluding first the monoclinic phase and second the orthorhombic phase. They concluded that the crystal phases in the films were monoclinic for the pure $HfO_2$ film, tetragonal or cubic for the Y(15%):$HfO_2$ film, and orthorhombic for the Y(7%):$HfO_2$ film. Therefore, the metastable orthorhombic phase was only stabilized in the epitaxial $Hf_{0.93}Y_{0.07}O_2$ film. Shimizu et al. performed also XRD measurements of this sample at high temperature (Figure 1a).[33] The weak asymmetric orthorhombic (o) o-YHO(110) reflection was measured from room temperature to 600 °C (Figure 1b). The peak vanished at around 450 °C, signaling a transition to paraelectric tetragonal or cubic phase, and thus indicating that the Curie temperature of the film was around 450 °C.

Orthorhombic films on cubic substrates are expected to present crystal variants. This was experimentally observed by XRD in YHO films on YSZ(001).[34] Local analysis by transmission electron microscopy (TEM) revealed a lateral size of the crystallites of around 10 nm, and that grain boundaries were along {100} or {110} planes.[35] Shimizu et al.[36] observed that films on indium-tin oxide (ITO) buffered YSZ(001), presenting only b- and c-oriented domains (c-axis was considered the polar axis), had large polarization of around 30 μC/cm$^2$. This suggested that ferroelectric switching should imply ferroelastic domain switching, which was later confirmed experimentally by micro-diffraction, focusing a synchrotron radiation beam on poled and pristine capacitors (Figure 2) where vanishing of (0l0) reflections was observed after junction cycling indicating a polarization rotation from in-plane to out of plane.

### 2.1. Control of film orientation on yttria-stabilized zirconia

Funakubo and collaborators prepared also epitaxial $Hf_{0.93}Y_{0.07}O_2$ films, around 15 nm thick, on ITO buffered YSZ(110)[37] and YSZ(111)[38] substrates. YHO films and ITO electrodes were epitaxial and presented cube-on-cube like epitaxial relationship with the substrate. Ferroelectric polarization loops were measured, being the remanent polarization around 10 and 16 $\mu C/cm^2$ for (111) and (110) oriented films, respectively. The coercive field $E_c$ of both films was similar, around 2 MV/cm.

XRD measurements at high temperature of a YHO(110) film[37] signaled $T_c$ around 450 °C, similar to YHO(001) films.[33] Recently, the same research group has extended this analysis to (111)-oriented epitaxial films and polycrystalline films, also characterizing films with different thickness.[39] Although the investigated thickness ranges and growth conditions were not coincident, the results suggest small orientation and thickness dependence on $T_c$ in films thicker than around 10 nm, with $T_c$ around 500 °C, and reduction of $T_c$ in thinner films, $t \sim 5$ nm, up to 350 °C (Figure 1c).

### 2.2. Other dopants and compositions

Li et al.[40] deposited $Hf_{0.5}Zr_{0.5}O_2$ (HZO) films, 15 and 50 nm thick, by PLD at 700 °C on TiN buffered (001), (110) and (111) oriented YSZ single crystals. TiN electrodes were (110)-oriented on YSZ(110), and (111)-oriented on YSZ(100) and YSZ(111). Epitaxy of HZO was confirmed by reflection high-energy electron diffraction and TEM. XRD characterization (symmetrical scans) showed peaks that were indexed as o-HZO(121) on TiN(110) and o-HZO(111) on TiN(111). However peaks positions were far to the nominal position of the corresponding reflections, which might indicate different phase. Nevertheless, polarization loops were measured for films on (001), (110) and (111) -oriented YSZ substrates, with remanent polarization in the 7-20 $\mu C/cm^2$ range and coercive field in the 1.1 - 2.3 MV/cm range.

Shiraishi et al. prepared epitaxial Fe-doped $HfO_2$[41] and Ce-doped $HfO_2$[42] films. Films were grown on ITO/YSZ(001) by ion-beam sputtering at room temperature followed by rapid thermal annealing. Fe-doped films presented the greatest amount of orthorhombic phase for a Fe content of around 6%. The polarization loops, not saturated, showed $P_r$ of around 8.8 $\mu C/cm^2$. The Curie temperature, estimated from XRD measurements at high-temperature, was around 500 °C. In the case of Ce-doped films, XRD and TEM characterization suggested that films doped with 3 to 10% Ce are orthorhombic. The polarization loops showed low remanent polarization, being the estimated value around 5 $\mu C/cm^2$.

## 3. Solid phase epitaxy

Kiguchi et al.[43] reported epitaxy of HZO by crystallization of amorphous films. A HZO film, deposited by ion-beam sputtering at room temperature, was crystalized by rapid thermal annealing (RTA) at 800 °C for 10 min under oxygen flow. The film was 30 nm thick. The authors concluded, from high-angle annular dark-field imaging (HAADF) scanning transmission electron microscopy (STEM) images compared with simulations, that the crystallized epitaxial film included both monoclinic and orthorhombic phases. However, the measured XRD scan showed a very broad peak that did not allow to confirm the stabilization of the orthorhombic phase. A similar growth process was done by Mimura et al.,[39] using PLD to grow non-ferroelectric Y-doped $HfO_2$ films at room temperature on YSZ(111) (Figure 3a) and subsequent RTA at around 1000 °C under $N_2$ atmosphere (Figure 3b). The ferroelectric properties were similar to those of films grown by PLD at 700 °C.[34, 37] Films around 15 nm present orthorhombic phase, while there is coexistence of orthorhombic and monoclinic phases in films around 30 nm thick. In-situ XRD during annealing of 15 nm thick $Hf_{0.93}Y_{0.07}O_2$ films on ITO/YSZ(111) allowed identifying the evolution of crystal phases during heating to 1000 °C and cooling to room temperature.[44] The film deposited at room temperature was monoclinic, and phase transformation was not detected up to 600 °C. Above this temperature monoclinic and tetragonal phases coexisted, being the film purely tetragonal above around 950 °C. Cooling the film, it remained tetragonal up to around 400 °C, when both tetragonal and orthorhombic coexisted. Finally, only the orthorhombic phase is detected below 200 °C (Figure 3c,d,e).

Suzuki et al.[45] investigated the influence of the atmosphere in sputtering at room temperature and rapid thermal annealing for crystallization. They deposited $Hf_{0.93}Y_{0.07}O_2$ films, around 24 nm thick, by RF magnetron sputtering under 200 mTorr of pure Ar or Ar/$O_2$ = 100. Crystallization was done at 1000 °C for 10 s under $N_2$ or $O_2$ flow. The orthorhombic phase was present in all samples, but the ferroelectric properties were very different. The film deposited in pure Ar and crystallized under $N_2$ flow showed well saturated polarization loops, while loops could not be measured in the films grown in Ar gas with around 1% $O_2$ due to leakage. The film deposited under pure Ar and annealed under $O_2$ showed ferroelectric loops, but with significant imprint and low breakdown field. Current leakage curves of films deposited with partial pressure of $O_2$ or crystallized under $O_2$ flow were asymmetrical and showed large leakage. Similar influence of oxygen is observed in polycrystalline films prepared by sputtering.[46-47]

Recently, Mimura et al.[48] reported the unexpected epitaxial growth of orthorhombic $Hf_{0.93}Y_{0.07}O_2$ films at room temperature without annealing. The films were deposited by sputtering in pure Ar atmosphere. The absence of $O_2$ in the sputtering gas was found to be critical in the formation of orthorhombic phase, and films deposited under Ar/$O_2$ = 100 were monoclinic. This result, which suggests that oxygen vacancies are convenient to stabilize the metastable orthorhombic phase, are in agreement with observations done with polycrystalline films[46-47] and epitaxial films obtained by RTA of amorphous films.[45] Mimura et al.[48] confirmed ferroelectricity by polarization loops, with high $P_r$ of 14.5 µC/cm$^2$ in films on ITO/YSZ(111), and much lower polarization in films on ITO/YSZ(001). Although it is known that some fluorite oxides as $CeO_2$ can grow epitaxially at room temperature,[49] the result reported by Mimura et al.[48] was surprising and further studies are required to understand the mechanism that allow epitaxial growth of the metastable phase at room temperature.

## 4. Epitaxial ferroelectric $HfO_2$ films on perovskite single crystals

Si has been widely used as a dopant to stabilize the ferroelectric phase in polycrystalline films. This is achieved for a narrow range of Si content of around 4 %, while for slightly higher content the tetragonal phase stabilizes and the films are antiferroelectric.[8, 24] The epitaxial growth of Si-doped $HfO_2$ was reported by Li et al.[50] They deposited Si(4.4%):$HfO_2$ films (Si:HO), with thickness in the 1.5-15 nm range, on (001), (110) and (111) oriented Nb-doped $SrTiO_3$ (STO) semiconducting substrates. Peaks in symmetric $\theta$-2$\theta$ XRD scans were indexed as the orthorhombic o-Si:HO(002) reflections in films on the three substrates, while additional peaks on Nb:STO(001) were indexed as the o-Si:HO(111) reflection. We note that peaks indexed as o-Si:HO(002) could also correspond to {200} reflections of the monoclinic phase. Similarly, in the case of the thicker films on Nb:STO(001), a second peak indexed as relaxed o-Si:HO(111) might also corresponds to monoclinic (m) m-Si:HO(-111). Polarization loops were reported clearly evidencing the ferroelectric character of the samples (Figure 4). The coercive fields were huge around 5 MV/cm and with little dependence on thickness and sizeable leakage contribution probably related to the use of Nb:STO as an electrode. Ferroelectric loops for 3 nm thick films were also reported without clear evidence of ferroelectric polarization.

An alternative to semiconducting Nb:STO perovskite substrates is the use of conducting epitaxial perovskite electrodes. Various groups have reported on epitaxial ferroelectric HZO films on $La_{0.67}Sr_{0.33}MnO_3$ (LSMO) electrodes grown epitaxially on STO(001) and other single crystalline perovskite substrates. Wei et al.[51] and Lyu et al.[52] reported ferroelectric HZO films on LSMO/STO(001) in 2018. The films were grown by PLD at 800 °C and 0.1 mbar of oxygen. The epitaxial films were (111) oriented, being the (111) diffraction peak in the $\theta$-$2\theta$ scans located at around 30° (Cu $K_\alpha$ radiation), while the peak is usually at around 30.5° in polycrystalline films. The films present four in-plane crystal variants (Figure 5a), as expected considering the 3-fold symmetry of HZO(111) and the 4-fold symmetry of LSMO(001). Wei et al.[51] deposited films, having thickness in the 4 - 27 nm range, and showed polarization loops for t = 9 nm and t = 5 nm films, with remanent polarization of 18 μC/cm$^2$ in the t = 9 nm and huge 34 μC/cm$^2$ in the t = 5 nm. There was not wake-up effect, although the films presented very strong fatigue and polarization was very low after $10^3$ cycles. The coercive fields were great, above 3 and 5 MV/cm in the t = 9 and 5 nm, respectively. On the other hand, $E_c$ was higher in the thinner film, as usually observed in ferroelectric perovskite oxides. Lyu et al.[52] compared two films, t = 9 and 18 nm thick, (111) oriented and with four in-plane crystal variants too. The presence of monoclinic phase coexisting with the polar orthorhombic phase was evident in XRD measurements of the thicker film using a 2D detector (Figure 5b). It was observed that the monoclinic phase has much higher mosaicity than the o-phase. Thus, the relative amount of monoclinic phase in HZO films on LSMO/STO(001) can be underestimated by XRD $\theta$-$2\theta$ scans using a point detector. The increase of monoclinic phase with thickness in epitaxial films is commonly observed too in polycrystalline HZO films.[14-16] Lyu et al.[52] reported ferroelectric loops, without wake-up effect (Figure 6a). The t = 9 nm film exhibited huge coercive field above 3 MV/cm and imprint field of 0.5 MV/cm directed towards the LSMO bottom electrode. The remanent polarization was around 20 μC/cm$^2$ in this film, and less than 7 μC/cm$^2$ in the t = 18 nm film. These polarization values, and those reported by Wei et al.[51] for t = 5 and 9 nm films, pointed to a decrease of polarization increasing HZO thickness, which could be connected with the larger amount of non-polar monoclinic phase.

Lyu et al.[52] demonstrated that epitaxial $HfO_2$ can present high endurance and retention. The t = 9 nm film was cycled for more than $10^8$ cycles without breakdown, although exhibited fatigue, i.e. polarization decreased from $2P_r$ = 25 to 9 μC/cm$^2$ after $10^8$ cycles of amplitude 4.5 V (Figure 6b,c). The dielectric permittivity showed too monotonic reduction with the number of cycles (Figure 6c, inset). Leakage current was constant during the first $10^4$ cycles, but it increased monotonously with additional cycles.[52] The retention was measured at room temperature after poling the film at 5.5 V up to $10^4$ s. Extrapolation of the experimental data pointed to retention well beyond 10 years.

### 4.1. Growth window for epitaxial stabilization

The growth window (substrate temperature and oxygen pressure) for epitaxial stabilization by PLD was reported by Lyu et al.[53] They grew t = 9 nm HZO films on LSMO/STO(001) varying substrate temperature in the $T_s$ = 650 - 825 °C range (at fixed $P_{O2}$ = 0.1 mbar) and varying oxygen pressure in the $P_{O2}$ = 0.01 - 0.2 mbar range (at fixed $T_s$ = 800 °C). The orthorhombic phase was formed in all the $T_s$ range, increasing the intensity of the o-HZO(111) peak with $T_s$, signaling increased amount of polar phase and/or reduced mosaicity. The out-of-plane lattice parameter, $d_{o\text{-}HZO(111)}$, decreased from around 2.98 Å to around 2.96 Å. All films were very flat, with root mean square roughness (rms) roughness less than 4 Å. The oxygen pressure has a great effect in the crystallinity of the films. Films deposited at 0.05 mbar or higher $P_{O2}$ exhibited an intense o-HZO(111) diffraction peak, while there were no diffraction peaks in the $P_{O2}$ = 0.01 mbar film. In the case of the $P_{O2}$ = 0.02 film there was a low intensity o-HZO(111) peak, and XRD measurements using a 2D detector revealed a coexisting high-mosaicity monoclinic phase. The amount of the monoclinic phase was lower in films deposited at higher $P_{O2}$. $d_{o\text{-}HZO(111)}$ decreased with $P_{O2}$, from close to 2.99 Å in the 0.02 mbar film to around 2.95 Å in the 0.2 mbar film. The rms roughness was less than 4 Å in films deposited at $P_{O2}$ up to 0.1 mbar, increasing in the 0.2 mbar film up to around 7 Å.

Substrate temperature and oxygen pressure have a notorious influence on the ferroelectric properties. Lyu et al.[53] reported that remanent polarization $P_r$ and coercive electric field $E_c$ of HZO films (t ~ 9 nm) increased with $T_s$, from 14 μC/cm$^2$ and 1.9 MV/cm to 21 μC/cm$^2$ and 2.9 MV/cm. $P_r$ increased also with $P_{O2}$, from negligible value in the 0.01 mbar film to around 20 μC/cm$^2$ in the 0.1 mbar film, being around 18 μC/cm$^2$ in films grown under higher $P_{O2}$. The dependence of $V_c$ on $P_{O2}$ was similar, with values in the 0.8 - 2.8 MV/cm range. Considering both $T_s$ and $P_{O2}$ series, there was clear correlation of polarization with amount of orthorhombic phase, while there was not an evident dependence with $d_{o\text{-}HZO(111)}$. Similar correlation between polarization and orthorhombic phase amount had been reported for polycrystalline doped $HfO_2$ films.[54] On the other hand, leakage current decreased more than one order of magnitude with $T_s$ (from ~ $10^{-5}$ A/cm$^2$ ($T_s$ = 650°C) to less than $10^{-6}$ A/cm$^2$ ($T_s$ = 825°C) at E = 1 MV/cm), while it increased more than three orders of magnitude with $P_{O2}$ (from ~ $10^{-7}$ A/cm$^2$ ($P_{O2}$ = 0.02 mbar) to more than $10^{-4}$ A/cm$^2$ ($P_{O2}$ = 0.2 mbar) at E = 1 MV/cm). The dependence on $P_{O2}$ suggests that leakage in these epitaxial films arises from other factors in addition to oxygen vacancies The leakage of HZO films exhibiting high polarization is around $10^{-6}$ A/cm$^2$ at E = 1 MV/cm, similar to other polycrystalline HZO films of equivalent thickness, although an extremely low leakage current of less than $10^{-8}$ A/cm$^2$ at E = 1 MV/cm has also been reported for t = 10 nm polycrystalline HZO films.[55]

## 4.2. Interfacial layer and epitaxy mechanisms

Wei et al.[51] characterized by cross-sectional STEM a HZO/LSMO/STO(001) sample and observed an interfacial HZO layer around 2-3 monolayers thick at the interface with LSMO. The layer, coherently strained, was identified as tetragonal phase. Later, Nukala et al.[56] determined by electron energy loss spectroscopy that the interfacial layer was oxygen deficient. They proposed that this oxygen deficient layer could be conducting, and that it could yield an additional screening mechanism for the stabilization of the polar phase. There was no interfacial layer in films on LSMO/$LaAlO_3$(001). Since it is known that the formation of oxygen vacancies is favored under tensile stress, the authors suggested that tensely strained LSMO was responsible for the formation of the tetragonal interfacial layer. Estandía et al.[57] also observed by STEM an interfacial layer in HZO/LSMO/STO(001). It was a pseudomorphic HZO layer, one monolayer thick, separated from the rest of the HZO film by a semicoherent interface. Similar interfacial layers were previously reported in interfaces between other highly mismatched oxides.[58]

Epitaxy generally happens by matching of atomic columns at the interface between two materials. Cube-on-cube epitaxy, or other simple epitaxial relationships as 45° in-plane rotation of the crystal lattices, are common. These conventional epitaxy habits are not viable in case of symmetry dissimilarity and high lattice mismatch, as it is the case of orthorhombic HZO(111) on (001) surfaces of LSMO or other perovskites of similar lattice parameter. However, unconventional mechanisms, as tilted epitaxy[59] or domain matching epitaxy (DME),[60-61] can permit epitaxy in some of these cases. Tilted epitaxy of HZO(111) occurs on Nb:STO(001), with a misorientation of about 15° with respect to the surface normal.[56] In the case of orthorhombic HZO(111) on LSMO(001), Estandía et al.[57] confirmed DME mechanism. The epitaxial relationships, determined by XRD, are [-211]HZO(111)//[110]LSMO(001) and [2-20]HZO(111)//[110]LSMO(001). The corresponding lattice mismatch ($f$ (%) = 100 x ($d_{LSMO}$-$d_{HZO}$)/$d_{HZO}$) in this heteroepitaxial system is huge: -9.97% along HZO[-211] and +58% along HZO[2-20]. The huge lattice mismatch discards conventional epitaxy of elastically strained growth and plastic relaxation above a critical thickness. In contrast, in DME m planes of the top layer nucleate on n planes of the bottom layer (m/n domain), with immediate formation of an extra or missing plane. This reduces the effective lattice mismatch $f^*$ = 100 x ($m \cdot d_{LSMO}$-$n \cdot d_{HZO}$)/$m \cdot d_{HZO}$) to a lower value. The coexistence of m/n and m'/n' domains allows an overall null mismatch. STEM characterization of the HZO/LSMO interface confirms it, with 9/10 and 10/11 domains in the [-211]HZO//[110]LSMO interface (Figure 7a,b) and 3/2 and 2/1 domains in the [2-20]HZO//[110]LSMO interface (Figure 7c,d).

### 4.3. Polar phase stabilized on LSMO

Wei et al.[51] observed that the out-of-plane $d_{(111)}$ lattice parameter was larger in the thinner films. They performed $\phi$-scans and $\theta$-2$\theta$ scans around asymmetric {111} reflections, using synchrotron radiation and measured $d_{(11-1)} = d_{(1-11)} = d_{(-111)}$ = 2.94 Å, smaller than the $d_{(111)}$ = 2.98 Å. The values were consistent with a rhombohedral unit cell, and the authors considered it a R3m rhombohedral metastable polymorph stabilized by the epitaxy. This was based on DFT calculations that simulated a R3m structure that was strongly polar (P ~ 15 µC/cm$^2$) under high epitaxial compressive stress causing an out-of-plane d(111) lattice parameter of 3.25 Å (Figure 8). Another polar rhombohedral phase, R3, was simulated for pure $HfO_2$, but it was less stable for HZO. Thus, it was concluded that ferroelectricity in epitaxial films on LSMO(001) was due to the polar R3m rhombohedral phase. More recent DFT calculations by Zhang et al. [62] confirmed high polarization in the R3m phase under huge compressive strain of around 5%, although the rhombohedral phase was only more stable than the orthorhombic $Pca2_1$ for extremely small thickness less than 5 five Hf-$O_2$ monolayers.

The DFT calculations [51] determined that the R3m phase should be under huge out-of-plane strain, $d_{(111)}$ > ~3.25 Å, to present the high polarization observed experimentally. Huge out-of-plane strain was experimental reported by Wei et al. [51], but only for an ultra-thin film of 1.5 nm. The polarization of that film was not reported, likely because the extremely low thickness did not allow the measurement. Experimental data of polarization of epitaxial HZO(111) on LSMO/STO(001) are reported for films of thicknesses ranging around 4.5 - 40 nm.[51-53, 63] The experimental $d_{(111)}$ out-of-plane lattice parameters of these films, ~2.96 - 3.05 Å, are much smaller than the very high values ($d_{(111)}$ > ~3.25 Å) that the DFT calculations found necessary for high polarization [51]. Besides, polarization shows moderate dependence on strain (and decreasing with $d_{111}$ instead of increasing) in films deposited on substrates with varying lattice parameter[64] or prepared varying growth condition (Figure 8).[53] Moreover, the $d_{(111)}$ extracted parameter is below 2.99 Å for any sample and thus far from the calculated necessary distortion to achieve large remanent polarization. Moreover, as shown below in Section 5, epitaxial HZO films on buffered Si(001), under great in-plane tensile stress when are cooled after growth due to the mismatch in thermal expansion coefficients between HZO and Si, present out-of-plane $d_{(111)}$ ~2.94 - 2.96 Å,[63, 65-66] significantly smaller than equivalent films on STO(001),[53, 63] but contrary to DFT predictions for R3m phase,[51] the polarization is larger. The experimental reports on evaluation of remanent polarization and $d_{(111)}$ out-of-plane lattice parameters similar to that found necessary for high polarization by DFT calculations[51] are non-existent. Wei et al.,[51] noted the possibility that some effects (crystallite size; surfaces and interfaces; local deviations from the average composition) could help to stabilize particular phases. Therefore, further studies are needed to conclude that the rhombohedral unit is a genuine phase, which could present ferroelectric properties different to those of the $Pca2_1$ orthorhombic phase, and not a simple structural distortion caused by epitaxial stress as commonly occur in other heteroepitaxial systems.

Materlik et al.[20] performed DFT calculations to compare the total energy of different polymorphs, including strain with an estimation of the surface energy. They concluded that strain was not the major cause of stabilization of the ferroelectric $Pca2_1$ phase, and that the stability of this phase in epitaxial HZO films was limited to size below 5 nm. Later, Liu et al.[67] considered epitaxial films of different orientation and strain and evaluated by DFT calculations the energy of two orthorhombic phases (the polar $Pca2_1$ and the non-polar Pbca) in comparison with the energy of the monoclinic $P2_1/c$. They concluded that the orthorhombic phases were more stable than the monoclinic phase in (111) films in a wide range of strain. The stable phase at high temperature would be tetragonal, and although the non-polar orthorhombic phase presented slightly lower energy than the polar phase, its transformation from the tetragonal phase was assumed to be slow and thus its formation could be suppressed by kinetics. Other recent DFT calculations evaluated epitaxial (111)-oriented films on STO(001) substrates and predicted that the phase responsible of the ferroelectricity was the polar orthorhombic phase $Pnm2_1$.[68] The authors proposed that the distinct $Pnm2_1$ phase would be the reason of the absence of wake-up effect in these epitaxial films. However, it has to be noted that wake-up effect is not a property genuine of a phase,[69-71] and different mechanisms can produce it.[3, 72-75] Assumptions of Qi et al.,[68] which can be extended to the other authors reporting DFT calculations of ferroelectric $HfO_2$ and other polymorphs, are based on strain states that differ to the experimental data revealed by XRD and STEM.[53, 57, 64]

#### 4.4. Epitaxial stress engineering

Estandía et al.[64] deposited LSMO electrodes and t = 9.5 nm HZO films on a set of ten (001)-oriented cubic or pseudo-cubic single crystalline oxide substrates ($YAlO_3$, $LaAlO_3$, $NdGaO_3$, LSAT, $SrTiO_3$ $DyScO_3$, $TbScO_3$, $GdScO_3$, $DyScO_3$, $NdScO_3$ and MgO), with lattice parameter ranging from 3.71 Å ($YAlO_3$) to 4.21 Å (MgO). The t = 25 nm LSMO electrodes grew coherently for lattice mismatch from around -2% (LSMO on $LaAlO_3$) to +2% (LSMO on $TbScO_3$), and were partially or fully relaxed on more mismatched substrates. XRD measurements using a 2D detector demonstrated that the in-plane parameter of LSMO determines the relative amount of orthorhombic and monoclinic phases in the HZO film. When LSMO ($a_{LSMO}$ = 3.874 Å in bulk) is compressively strained ($a_{LSMO}$ smaller than around 3.86 Å), HZO crystallizes in monoclinic phase. When LSMO is relaxed or under tensile strain, the metastable orthorhombic phase is stabilized, increasing the o/m ratio with $a_{LSMO}$. LSMO electrodes on scandate substrates were highly tensely strained ($a_{LSMO}$ ~3.95 Å), and the HZO films were almost purely orthorhombic. STEM characterization allowed to confirm the increased amount of orthorhombic phase for increasing substrate lattice parameter (Figure 9a,b,c). The films on LSMO/LSAT, LSMO/STO and LSMO/$GdScO_3$ showed single monoclinic phase, coexisting monoclinic and orthorhombic phases, and single orthorhombic phase, respectively. It is noted that grains are observed even in single phase films, due to the crystal variants existing in both monoclinic and orthorhombic phases. The grain boundaries between crystal variants are coherent, while those between monoclinic and orthorhombic phases are incoherent.[76]

Estandía et al.[64] reported polarization loops of the films deposited on the ten oxide substrates. The ferroelectric polarization was very low in the film on substrates ($YAlO_3$, $LaAlO_3$ and $NdGaO_3$) causing compressive strain to LSMO, and it was low ($P_r$ ~ 5 μC/cm$^2$) in the film on LSAT, on which LSMO is practically unstrained. Polarization increases in films on substrates with larger lattice parameter, being $P_r$ ~ 24 μC/cm$^2$ on $TbScO_3$. LSMO relaxes plastically on substrates with lattice parameter greater than $TbScO_3$, and the polarization of the HZO film on these substrates is lower. In the case of ferroelectric perovskites, polarization usually depends on the lattice strain of the ferroelectric film. It allows the control of the ferroelectric polarization by selection of a substrate that causes a certain elastic strain. This is usually called strain engineering.[77] However, in the case of epitaxial HZO films, the great influence of the substrate on the ferroelectric polarization is not mainly due to HZO strain (i.e., strain engineering). Figure 9d shows that $d_{o\text{-}HZO(111)}$ does not vary significantly in the HZO films and this small variation does not produce a significant $P_r$ variation. In fact, the small decrease of $P_r$ with $d_{o\text{-}HZO(111)}$ appreciated in Figure 9d is against the increase of polarization with $d_{o\text{-}HZO(111)}$ theoretically predicted and discussed in the previous section. Thus, other contributions play a more important role. In this series of HZO films on ten substrates, Estandía et al.[64] observed that films on scandate substrate had the greater amount of orthorhombic and the larger ferroelectric polarization (Figure 9e).

Therefore, in epitaxial growth of HZO/LSMO on oxide substrates, the selection of scandate substrates causing tensile strain is critical to have the highest ferroelectric polarization. The different substrates cause different strain in LSMO, which determines the relative amount of monoclinic and orthorhombic phases. Therefore, the impact on the properties of HZO is due to epitaxial stress engineering, but not to epitaxial HZO strain engineering.

Nukala et al.[56] deposited t = 6 nm HZO films on a set of six different oxide substrates ($YAlO_3$, $LaAlO_3$, $NdGaO_3$, LSAT, $SrTiO_3$ and $DyScO_3$) buffered with a t = 40 nm LSMO electrode. The HZO films on the two substrates with larger lattice parameter, STO and DSO, presented single polar rhombohedral phase, (111)-oriented. This phase was present too in films on LSAT and $LaAlO_3$, substrates with intermediate lattice parameters, but it was accompanied of (001)-oriented crystallites, attributed to mixture of monoclinic and tetragonal phases. These results confirmed that epitaxial stress engineering,[64] allows controlling the stabilization of the ferroelectric phase in HZO films enough thin: tensely strained LSMO electrodes favors it, while compressively strained LSMO suppresses it.

Epitaxial stress engineering is thus limited to substrates having moderately high lattice mismatch with LSMO, to allow its coherent growth. In addition to lattice mismatch, LSMO coherence or relaxation can depend on LSMO deposition parameters and thickness. Nukala et al.[56] grew HZO films on $LaAlO_3$(001) substrates buffered with t = 10 and 40 nm LSMO electrodes. HZO on the fully strained t = 10 nm electrodes, it presented only non-polar phases, while there was mixture of non-polar and polar phases in the HZO film on the partially relaxed t = 40 nm LSMO. However, Yoong et al.[78] reported divergent results for a t = 10 nm HZO film on fully strained 10 nm LSMO on $LaAlO_3$(001). From XRD measurements around asymmetrical reflections and STEM characterizations, the authors confirmed epitaxy in spite of the low growth temperature of 550 °C, and concluded that the film presented single polar orthorhombic phase, being (00l) oriented. The low deposition temperature could maybe favor the stabilization of the orthorhombic phase and its (001)-orientation under compressive stress, although other studies have shown that crystallization of films on STO(001) is inhibited decreasing substrate temperature below to around 650 °C.[53] Polarization loops confirmed that the (001)-oriented phase was polar, measuring remanent polarization around 20 $\mu C/cm^2$, with coercive fields of -2 and + 2.8 MV/cm.

**4.5 Impact of bottom electrode**

Epitaxy of ferroelectric $HfO_2$ has been reported using only a few bottom electrodes. ITO is the bottom electrode that is commonly used to grow epitaxial Y-doped $HfO_2$,[26, 36-38, 44-45, 48, 79-80] Ce-doped $HfO_2$,[42] and Fe-doped $HfO_2$[41] films. Semiconducting Nb:STO substrates were used to deposit epitaxial Si-doped $HfO_2$ films.[50] In the case of HZO, the epitaxial stabilization of the ferroelectric phase is achieved using a LSMO bottom electrode.[51-53, 56-57, 63-66, 76, 78, 81-89] Very recently, Estandía et al.[89] investigated the influence of the bottom electrode on the stabilized phases and the ferroelectric polarization of HZO films. They compared films on LSMO/STO(001) with others on conducting $LaNiO_3$ (LNO), $SrRuO_3$ (SRO), LSMO, and $Ba_{0.95}La_{0.05}SnO_3$ (BLSO) electrodes grown on STO(001), and also with a HZO film on a doped Nb:STO(001) semiconducting substrate. The orthorhombic phase does not form on LNO and SRO, but on Nb:STO and BLSO, showing an unusual tilted epitaxy. There was not ferroelectric polarization in the films on LNO and SRO, while in the films on Nb:STO and BLSO it was much smaller than that of films on LSMO. It was ruled out that the impact of the electrode was due to the epitaxial stress effects described in previous section. To get insight on the cause of the impact, LSMO electrodes covered with ~3 unit cells (u.c.) of LNO or SRO were used. The formation of the orthorhombic phase was suppressed. In contrast, when LNO or SRO electrodes were covered with ~3 u.c. of LSMO, the orthorhombic phase was epitaxially stabilized and the films showed high polarization. The results pointed that the different chemical composition and atomic structure of the surface of the electrodes was a relevant factor. Finally, Estandía et al.[89] compared four $La_{1-x}A_xMnO_3$ (A= Sr, Ca; x= 0.33, 0.5) electrodes, and it was found that the amount of the orthorhombic phase did not depend on the divalent atom (Sr or Ca), but it was greater for $La_{0.67}A_{0.33}MnO_3$ than for $La_{0.5}A_{0.5}MnO_3$. Taken together, the results pointed to a complex interface, critically influencing the epitaxial stabilization. Detailed studies of the atomic structure are necessary to determine the causes.

## 5. Epitaxial integration on Si

Lee et al.[90] deposited Y-doped $HfO_2$ films (thickness and Y content were not indicated) on YSZ/Si(001) by PLD. Symmetric $\theta$-$2\theta$ XRD scans (Cu K$\alpha$ radiation) showed a peak at around 35°, indexed as (001)/(010) reflections of the orthorhombic phase. This diffraction peak can also be ascribed to the YSZ(002) peak, as well as {200} peaks of the monoclinic phase. The authors also measured $\phi$-scans and reciprocal space maps around asymmetrical reflections, but overlapping with other reflections is also expected. Films were characterized electrically and, although the absence of bottom electrode, their ferroelectric nature was evidenced in spite of important leakage current contributions. $Hf_{0.93}Y_{0.07}O_2$ films, as thick as 900 nm, were also integrated epitaxially on ITO/YSZ/Si(001) by deposition at room temperature and RTA.[91] The films presented properties similar to the films of similar thickness on YSZ(001) with $P_r$ around 4 μC/cm$^2$ and coercive electric field of 1 MV/cm. Other authors deposited HZO films on bare YSZ(001) single crystals and reported the formation of monoclinic phase, without presence of orthorhombic phase.[92]

L. Bégon-Lours et al.[93] deposited epitaxial HZO films on GaN/Si(111) and reported structural characterization. The films were (111) oriented and from measurements of $d_{\{111\}}$ lattice distances, the unit cell was found to be rhombohedral. The films were characterized by STEM, and the authors suggested that the R3 rhombohedral phase matched better than R3m.

Nukala et al.[82] deposited HZO (x = 0.5, 0.7 and 0.85) directly on Si(111), and a x = 0.7 film was also grown on Si(001). Zr atoms likely reduced the native amorphous $SiO_x$, allowing epitaxy as occurs in epitaxial growth of yttria-stabilized zirconia.[94] They deposited epitaxial films by PLD, at $T_s$ = 800 °C under 0.005 mbar of Ar. From XRD and STEM characterization, the authors concluded that the phase of the film on Si(001) was orthorhombic, while the films on Si(111) presented mixture of monoclinic and rhombohedral phases, the latter more abundant in the HZO films with higher Zr content. I-V loops measured at 5 K evidenced ferroelectric character of films on Si(111) although the presence of large leakage currents.

$SrTiO_3$ is a natural option as buffer layer for epitaxial integration of ferroelectric $HfO_2$ on Si(001). On one hand, STO(001) substrates have been used extensively to grow ferroelectric HZO film. On the other hand, STO can be grown epitaxially on Si(001) by molecular beam epitaxy (MBE)[95] and other techniques as PLD[96] or atomic layer deposition.[97] Si(001) coated with a MBE grown STO layer was used to deposit by PLD a LSMO electrode and a t = 7.7 nm HZO film,[65] using same deposition parameters that in equivalent films on STO(001).[52] XRD confirmed epitaxial stabilization of the orthorhombic phase, with diffraction spots of the monoclinic phase barely observable. The out-of-plane interplanar spacing, $d_{o\text{-}HZO(111)}$ = 2.940 Å, smaller than that of the equivalent HZO films on LSMO/STO(001), was probably a consequence of the thermal expansion mismatch between the oxide layers and the Si(001) substrate. Topographic AFM images showed a flat surface with rms roughness of 3.5 Å. Polarization loops confirmed very large remanent polarization of 34 μC/cm$^2$, significantly larger than the 24 μC/cm$^2$ of films t $\sim$ 9 nm thick on STO(001) perovskite substrates.[53] The dielectric constant showed an hysteretic loop, with the butterfly shape characteristic of a ferroelectric, and permittivity at saturation of 31-32. The film presented excellent endurance behavior up to $10^9$ cycles of 4.0 V amplitude, although with fatigue that reduced $2P_r$ from ~30 μC/cm$^2$ in the pristine state to 4.5 μC/cm$^2$ after the $10^9$ cycles.

Remarkably, capacitors poled with the same voltage of 4.0 V retain ferroelectric polarization for more than 10 years.

YSZ, contrary to STO, can be (relatively) easily grown epitaxially on Si(001) by PLD.[94] For this reason, YSZ is probably the most popular buffer layer for epitaxial growth of functional oxides on Si(001). However, HZO/LSMO bilayers cannot be directly grown on YSZ/Si(001) due to the large lattice mismatch between LSMO and YSZ. This is solved by inserting oxide films to accommodate progressively the mismatch, and HZO/LSMO bilayers were grown epitaxially on $LaNiO_3/CeO_2$/YSZ/Si(001).[65] It is noted that the top buffer layer, $LaNiO_3$, is a conducting oxide with lattice parameter (3.84 Å) close to that of LSMO (3.875 Å). However, HZO films directly grown on $LaNiO_3$ were not orthorhombic .[65] This points to the critical role of the LSMO(001) surface on stabilization of the orthorhombic phase. Lyu et al.[85] reported the dependence on thickness in the t = 4.6 - 18.6 nm range. XRD measurements confirmed the epitaxial stabilization of the orthorhombic phase, and symmetric $\theta$-$2\theta$ scans showed the o-HZO(111) reflection in all the samples. XRD $2\theta$-$\chi$ frames were measured to detect the monoclinic phase. The spots corresponding to this phase increased strongly in intensity with thickness. All HZO films showed ferroelectric polarization hysteresis loops, with remanent polarization decreasing with thickness. The dielectric constant also showed hysteresis. The hysteresis decreases with thickness, and the permittivity at saturation reduces from ~33.5 to ~28. The observed increase of paraelectric monoclinic phase in thicker films is probably the cause of loss of polarization and decrease in dielectric constant. Figure 10a shows that remanent polarization decreases monotonically with thickness from ~33 $\mu C/cm^2$ (t = 4.6 nm) to ~10 $\mu C/cm^2$ (t = 18.6 nm). The monotonic dependence contrast with the usual peaky dependence reported for polycrystalline HZO films (empty symbols in Figure 10a). The increase in polarization with decreasing thickness of the epitaxial HZO films on Si(001) suggests that ferroelectricity could be maintained in extremely thin films. This is in agreement with theoretical calculations.[98] This was recently confirmed by Cheema et al.,[99] measuring ultrathin (t = 1 nm) $Hf_{0.8}Zr_{0.2}O_2$ polycrystalline films grown directly on Si(001) by piezoresponse force microscopy and demonstrating switchable polarization.

Remarkably, epitaxial HZO films on Si(001) show a higher remanent polarization than equivalent films on STO(001) substrates. It has been shown in Section 4.4, for HZO/LSMO grown on different oxide substrates, that the selection of $GdScO_3$ or $TbScO_3$ as substrate is critical to obtain the highest ferroelectric polarization. This was correlated with the tensile epitaxial stress that the scandate substrates cause on HZO/LSMO. In the case of the films on Si(001) substrates, the stress was not due to mismatch between lattice parameters, but to the lower thermal expansion coefficient of Si(001) that caused the compression of the out-of-plane parameter of orthorhombic HZO when the film was cooled after deposition.

## 6. Dependence of $P_r$ and $E_c$ on thickness

In polycrystalline doped-$HfO_2$ films, the relative amount of paraelectric monoclinic phase respect the ferroelectric orthorhombic phase is found to increase with film thickness.[14-16] This dependence, which can influence critically the ferroelectric properties, has been observed too in epitaxial YHO[80] and HZO films[53, 85] films.

Mimura et al.[80] prepared epitaxial YHO films of thickness in the 10-115 nm range on ITO/YSZ(111). The coercive field, similarly as it occurs in polycrystalline films of doped $HfO_2$,[100] did not change significantly with thickness. The polarization of the films showed small dependence on thickness. More recently, Shimura et al.[91] prepared $Hf_{0.93}Y_{0.07}O_2$ films as thick as 340 and 1080 nm. The films were deposited at room temperature by RF magnetron sputtering on ITO/YSZ(001) and ITO/YSZ(111), followed by RTA (800 °C, 10 s, $N_2$). Epitaxy was confirmed by XRD ($2\theta$-$\chi$ frames recorded with a 2D detector and pole figures). Films on YSZ(001) were epitaxial and single (001) oriented, although the {100} diffraction spots in the t = 1.08 µm were highly elongated along $\chi$, signaling high mosaicity. Films on YSZ(111) presented coexistence of (111) and (001) orientations, and the diffraction spots corresponding to both orientations were very elongated for both thicknesses. Polarization loops confirmed ferroelectricity in all samples, with constant remanent polarization (5 μC/cm$^2$) and coercive electric field (1 MV/cm) irrespective of film thickness and orientation (Figure 10b-c). The same group reported also high ferroelectric polarization (~15 μC/cm$^2$) in polycrystalline $Hf_{0.93}Y_{0.07}O_2$ films of around 1 μm in thickness prepared on Si(001) by PLD at room temperature and subsequent RTA.[44] These results contrast with the common observation of degradation of ferroelectricity increasing thickness of doped $HfO_2$ films. Usually, either polycrystalline[16] or epitaxial[53, 63, 85] doped-$HfO_2$ films show low or negligible polarization for thickness above a few tens of nanometers. On the other hand, the $E_c$-t dependence of the YHO epitaxial films (Figure 10c), and polycrystalline films of doped $HfO_2$ films,[48, 100] differ to the behavior of conventional ferroelectric perovskites. In high quality ferroelectric perovskite films, $E_c$ decreases with thickness, very usually following the $E_c \propto t^{-2/3}$ scaling law.[101-103] Thus, the preservation of ferroelectricity in polycrystalline[44] and epitaxial[91] $Hf_{0.93}Y_{0.07}O_2$ films could be likely consequence of the particular microstructure of films deposited at room temperature and crystallized by RTA. However, ferroelectricity in bulk crystals of Y-doped $HfO_2$ has been reported very recently,[104] demonstrating that this property is not restricted to nanometric sizes. The orthorhombic phase was kinetically stabilized in crystals that were quenched very rapidly. The ferroelectric polycrystalline[44] and epitaxial[91] ~1 μm thick $Hf_{0.93}Y_{0.07}O_2$ films were crystallized by a RTA process. However, polycrystalline $HfO_2$ films, doped with Y or other dopants, prepared by similar RTA processes show usually severe degradation of polarization when they are thicker than a few nanometers. Moreover, similar degradation is observed in films grown epitaxially in-situ and cooled much slowly. It remains to be determined the microstructural characteristics (oxygen vacancies and other defects, residual stress, etc.) that allow preservation of ferroelectricity in thick films.

In the case of epitaxial HZO films, Lyu et al.[53] reported critical impact of the thickness on the ferroelectric properties. A series of HZO films, with thickness in the 2.3 - 37 nm range, on LSMO/STO(001) was characterized. Increasing thickness, the amount of non-polar monoclinic phase increases. Films thinner than around 14 nm were very flat, with rms roughness less than 4 Å. In thicker films rms increased up to around 8 Å in the t = 37 nm film. Polarization loops were measured for films thicker than 4.6 nm (leakage of thinner films impeded ferroelectric measurements). The remanent polarization showed a peaky dependence on thickness, with the largest polarization of 24 μC/cm$^2$ for films around 7 nm thick (Figure 10a). Similar dependence is reported for polycrystalline films (Figure 10a[15, 100, 105-107]). The coercive field of the loops varies with thickness. The log-log plot of $E_c$ against thickness (Figure 10d) shows linear dependence with slope -0.61, in agreement with the $E_c \propto t^{-2/3}$ scaling law. The same dependence of coercive field on thickness was observed for epitaxial HZO films on Si(001)[85] and La-doped HZO films on STO(001) and Si(001).[63] It signals that crystal grain size along the out-of-plane direction is not smaller than the film thickness. In contrast, $E_c$ does not change significantly with thickness in epitaxial YHO films obtained by annealing of amorphous layers,[80, 91] or usually in polycrystalline $HfO_2$ films,[100] but yes (showing $E_c \propto t^{-2/3}$) for polycrystalline films thinner than around 30 nm and grains spanning across the entire thickness.[108]

## 7. Endurance and retention

A ferroelectric used in a memory has to show, in addition of high polarization, high retention and endurance. Endurance is perhaps the main limitation of ferroelectric hafnia for its future use in commercial devices. Endurance can be limited by hard breakdown (device failure) or by fatigue (polarization decrease upon cycling up to an undetectable value). Significant progress achieved in the last few years has permitted achieving endurance up to $10^{11}$ cycles,[85, 109] which is promising but still far of the $10^{15}$ cycles achieved with perovskites.[2] Further improving of endurance is also challenged due to the polarization - endurance dilemma: when a capacitor is optimized to enhance its endurance, this is accompanied of a reduction of the remanent polarization.[3] Moreover, it is convenient for memory reliability that polarization remains as constant as possible during its operational use. However, both wake-up effect and fatigue cause that polarization varies with number of cycles.

The first measurements of endurance in epitaxial HZO films revealed that fatigue was important.[52] Fatigue and retention can depend on thin film deposition parameters, and thus these properties have to be considered to determine the growth window of the thin films.[84] Fatigue and retention were measured in the series of films on LSMO/STO(001) deposited at varied $T_s$ and $P_{O2}$, and which polarization, coercive field and current leakage were previously reported.[53] To quantify fatigue, films were cycled $10^7$ times. Tiny wake-up effect was only observed in the films deposited at 700 °C or lower temperature, and there was absence of wake-up in the other films. This confirmed that epitaxial films are more robust against wake-up than polycrystalline films. In contrast, all the epitaxial films suffered fatigue (Table II). The relative polarization decrease upon cycling with respect the initial $P_r$ value increases with either $T_s$ or $P_{O2}$. In the case of the $T_s$ series, the remanent polarization of the 650 °C (825 °C) sample was 47% (27%) after $10^7$ cycles. In the case of the $P_{O2}$ series, the remanent polarization of the 0.02 mbar (0.15 mbar) sample was 42% (25%) after $10^7$ cycles. The 0.2 mbar sample was more fatigued before hard breakdown after only $10^3$ cycles. On the other hand, retention measurements showed asymmetry, being more stable the capacitor poled applying positive voltage to the top Pt electrode. The asymmetry was very high in the films deposited at low $T_s$, while asymmetry was moderately low at high $T_s$ independently of $P_{O2}$. Retention was short, for negative poling, in the films with combined high asymmetry and low coercive field. Excluding these films, polarization was very stable, with more than 40% of the initial value retained after 10 years at room temperature. Retention and fatigue made the growth window of the epitaxial films narrower than considering only polarization, but optimal $T_s$ and $P_{O2}$ ranges are still sufficiently wide. Figure 11 summarizes the effects of $T_s$ and $P_{O2}$ on polarization, retention and fatigue obtaining the best performance for $T_s$ around 750 °C and $P_{O2}$ = 0.10-0.15 mbar.

Lyu et al.[85] reported endurance and retention properties in epitaxial HZO films of thickness in the t = 4.6 - 18.4 nm range, deposited on LSMO/$LaNiO_3$/$CeO_2$/YSZ buffered Si(001). The retention (Figure 12a) after 10 years was extrapolated by fitting the remanent polarization measured at room temperature as a function of delay time $t_d$ to the $P_r = P_o t_d^{-k}$ dependence. All films presented net ferroelectric polarization after ten years, being retention longer in the thinnest films. It was argued[85] that the smaller fraction of monoclinic phase in these films could reduce the depolarizing field. Endurance measurements showed only tiny wake-up effect in the thinnest film. However, all samples presented fatigue. The loss of normalized polarization of the epitaxial HZO films on Si(001) with number of cycles did not depend on the electric field.[65] The thinner films presented significantly higher endurance than the thicker ones[85] and the sub-5 nm film showed excellent endurance ($2P_r$ was 6 μC/cm$^2$ after $10^{11}$ cycles of 2.5 V amplitude). Applying higher poling voltage, the initial polarization was higher, but hard breakdown occurred after a number of cycles (smaller as higher the voltage was). The authors measured permittivity hysteresis loops and current leakage during the endurance tests. The permittivity reduced gradually, similarly as observed in polycrystalline capacitors.[73, 110] Current leakage did not vary significantly initially, but it increased suddenly after high number cycles. The threshold occurred earlier as higher was the electric field, and it was not observed for low enough electric field. A gradual reduction of the imprint field was also observed during the endurance measurements. Evolution in the charge defects, mainly oxygen vacancies, were proposed to cause the observed variation in polarization, imprint and leakage, as suggested for polycrystalline hafnia films.[73]

Figure 12b compares the polarization and endurance of the epitaxial sub-5 nm HZO film on Si(001) with data reported for polycrystalline films. The combination of high endurance and polarization in the epitaxial films equals the best results reported for polycrystalline films, which was achieved doping HZO with La.[109] A dilemma between endurance and retention[111-113] could constitute a bottleneck for further reliability improvement. Retention and endurance are generally evaluated considering different samples or different poling fields, while both properties should be optimal under same operation conditions. The epitaxial films reported by Lyu et al.[85] presented endurance of at least $10^{11}$ cycles and retention longer than 10 years using same poling field. The thickness of the films, less than 5 nm, can allow its use in tunneling devices.

Adkins et al.[114] measured endurance of an epitaxial HZO film, t = 5.6 nm, on LSMO/$LaNiO_3$/$CeO_2$/YSZ/Si(001) at cryogenic temperatures. The fatigue present at room temperature diminished reducing temperature, and there was not loss of polarization at 33 K in the explored range of $10^7$ cycles, even for huge applied field of around 10 MV/cm. The authors estimated an activation energy of around 23.4 meV associated with fatigue. The value appeared to be too low for oxygen vacancies diffusion, and diffusion of trapped injected charges was proposed as main cause of fatigue.

In polycrystalline HZO films, La doping causes reduction of the coercive field and leakage, allowing endurance enhancement.[109, 115] Epitaxial 1% mol La doped HZO films, with thickness in the 4.5 - 13 nm range, were grown by PLD on LSMO/STO(001) and LSMO/STO/Si(001).[63] The coercive field was not reduced respect to equivalent epitaxial undoped HZO films, although leakage diminished by around one order of magnitude (leakage of t = 8.3 nm films was $\sim 10^{-7}$ A/$cm^2$ at 1 MV/cm). The remanent polarization was higher than in equivalent undoped epitaxial HZO films, particularly in films thicker than around 10 nm, probably due to lower amount of monoclinic phase. The epitaxial La-doped films thinner than around 8 nm exhibited wake-up effect, although only for around 100 cycles, while in polycrystalline La-doped films it can occur up to $10^7$ cycles.[111] Endurance in the epitaxial films was limited by fatigue or hard breakdown, the latter occurring earlier as higher the cycling field is. A 4.6 nm thick La-doped HZO film on LSMO/STO(001) showed the highest endurance, without breakdown after $5 \times 10^{10}$ cycles of 5.4 MV/cm, but with $2P_r$ reduced to around 3 μC/$cm^2$ due to fatigue. Films thicker than 5 nm presented long retention, with high remanent polarization extrapolated after 10 years at room temperature. This was measured after poling at different voltages, including the low voltages that allowed high endurance. Also, retention longer than 10 years was measured at 85 °C of t = 8.3 nm films on LSMO/STO(001) and LSMO/STO/Si(001). The results demonstrated that wake-up effect and endurance-retention dilemma are not intrinsic in La-doped HZO. Very recently, it has been reported on the epitaxial stabilization of the orthorhombic phase in La-doped $HfO_2$ films.[88] Films t = 12 nm thick on STO and $DyScO_3$ substrates showed remanent polarization of around 16 μC/$cm^2$, while t = 8 nm films were very leaky and t = 16 nm films showed smaller polarization. The retention was more than 10 years and the endurance about $10^7$ cycles. The authors investigated in detail the influence of pulse voltage amplitude and width, and interval between pulses, on the endurance, observing a big impact on wake-up and fatigue. Contrary to the observations that fatigue does not depend on pulse amplitude in epitaxial HZO[65, 85] and La-doped HZO films,[63] fatigue in La:$HfO_2$ depended strongly on the applied electric field, being more severe for the applied voltage and pulse width smaller. Li et al.[88] considered domain wall pinning as the main mechanism of fatigue, and that high amount of domain walls in case of partial switching accelerated fatigue.

Measurements of the endurance of epitaxial doped $HfO_2$ films confirm a different behavior from that of polycrystalline films. Polycrystalline films exhibit a recurrent wake-up effect, even up to around $10^7$ cycles, but can be fatigue-free.[109] The wake-up effect in epitaxial films on LSMO electrodes is negligible, or limited to a few cycles in ultra-thin films.[63, 116] LSMO is not expected to induce formation of oxygen vacancies in the doped $HfO_2$ film, unlike the TiN electrodes that are commonly used in polycrystalline capacitors. This can be a major factor in the near absence of a wake-up effect in epitaxial films. In contrast, fatigue is a serious problem and epitaxial films without fatigue have not yet been reported. Wake-up and fatigue have been extensively investigated in polycrystalline films, and different mechanisms have been reported for both phenomena, including domain depinning/pinning at oxygen vacancies or other charged defects and paraelectric - ferroelectric phase changes.[73, 117-118] The multiple mechanisms observed experimentally suggest that the microstructure of the film may be critical. Paraelectric monoclinic and ferroelectric orthorhombic phases usually coexist in epitaxial films. The dependence of the fatigue of HZO films on LSMO/STO(001) with the deposition conditions ($T_s$ and $P_{O2}$), shown in Figure 11, reveals less fatigue in films grown at low $T_s$ or low $P_{O2}$, despite the fact that these films presented the highest amount of monoclinic phase, i.e less initial remanent polarization . It should be noted that the parasitic phase can promote fatigue through domain pinning, but could also suppress the propagation of pinned domains. Measurement of the endurance of epitaxial films on LSMO buffered scandate substrates, almost free of paraelectric phase,[64] would give insight on this.

Table I summarizes the ferroelectric properties of epitaxial films of doped $HfO_2$. Among the chemical compositions investigated, the highest remanent polarization of around 30-34 $\mu C/cm^2$ is reported for $Hf_{0.5}Zr_{0.5}O_2$,[51, 66, 116] La-doped $Hf_{0.5}Zr_{0.5}O_2$[63] and Si-doped $HfO_2$ films.[50] High endurance of up to $10^{11}$ and $5x10^{10}$ cycles has been measured for $Hf_{0.5}Zr_{0.5}O_2$[116] and La-doped $Hf_{0.5}Zr_{0.5}O_2$[63] films, respectively. Retention of more than 10 years was shown using writing voltage pulses of the same amplitude to those used to measure the endurance. Therefore, epitaxial doped $HfO_2$ films simultaneously allow high polarization, endurance and retention.

**8. Devices**

Ferroelectric $HfO_2$ is very interesting for memory devices and it has already used to fabricate 3D Fe-RAMs[10] and Fe-FETs with 22 nm node.[11] However, the development of Fe-TJs has been less successful, and very low endurance was reported. [119] Epitaxial films, with lower surface roughness and defects than polycrystalline films, single orientation and possibility of control of parasitic polymorphs, may offer opportunities to develop these challenging devices.

Electroresistance (ER) loops were soon measured in epitaxial HZO films a few nanometers thick.[76, 81, 83, 87] Sulzbach et al.[87] reported electroresistance loops with the same coercive voltage (Figure 13a) than the corresponding polarization loops (Figure 13b), indicating the close relation between ferroelectric switching and electroresistance. The Brinkman model was used to fit I-V characteristics pointing to tunneling current as predominant conduction mechanism. However, it was found that whereas short writing times or few cycling voltages result in below 1 order of magnitude electroresistance, the use of longer writing pulses or further sample cycling results in irreversible electroresistance. Thus, genuine tunneling ER resulting from ferroelectric switching can be accompanied by ER resulting from other transport mechanisms. They are associated to ions motion, and cause huge ER values up to ~$10^5$ %[87] or even ~$10^6$ %.[83] Noheda and coworkers[83, 86] proposed that the huge ER is connected to oxygen vacancy migration between bottom LSMO electrode and HZO film. However, other mechanisms can be considered too, as the widely studied filamentary resistive RAM devices based in Hf and Zr oxides suggest.[2, 120] It is understood that in these devices different resistance states are stabilized, because the formation of conducting channels, which are confined at the grain boundaries.[121-123] Therefore, filamentary ionic conduction can also play an important role in the found irreversible large electroresistance. Sulzbach et al.[76, 87] have followed two strategies to overcome this effect. The first is based in the control of the amount of orthorhombic phase by epitaxial stress.[64] It was observed that ionic conduction mechanism is less active (higher voltage is necessary to activate it) in HZO films grown on scandate substrates where predominant orthorhombic phase is found.[76] This points to the fact that monoclinic-monoclinic and orthorhombic-monoclinic, but not orthorhombic-orthorhombic, grain boundaries can favor ionic motion.[76] Second, conduction at the grain boundaries was also claimed to be reduced by covering the HZO layer with an insulating paraelectric layer.[87] The insulting layer blocks the ionic conduction channels resulting in tunneling electroresistance driven by ferroelectric polarization almost completely avoiding ionic conduction. Thus, as in polycrystalline films, important coexistence between ionic and ferroelectric contributions to electroresistance exists; however, epitaxial films have been proved to be unique to disentangle the FE and non-FE contributions to tunnel electroresistance.[76, 87]

The effect of the ionic exchange between the electrodes and the HZO has been used to manipulate not only the tunneling current, but also the tunneling magnetoresistance in FM/HZO/FM tunnel junctions, where FM layers act as electrodes.[81, 83] In Figure 13c,[83] it can be observed that the TMR inverts its sign for cycled samples, where ionic contributions are relevant. More relevant from applications point of view is that TMR is also modulated for samples where less ionic contribution is expected, thus indicating an intrinsic, but low, magnetoelectric coupling between the HZO and FM electrodes Figure 13d.[81]

Electroresistance is not only present in tunnel junction of ferroelectric materials. Thicker films can show conductivity due to thermionic injection at the interface which can be modulated by ferroelectric polarization.[124] Yoong et al.,[78] demonstrated this effect in 10 nm epitaxial HZO films. As expected in thick films, the electroresistance is much larger than in tunnel juntions. Interestingly, brain-like memristive activity was demonstrate with a throughout characterization. In Figure 13e,f it is shown the asymmetric Hebbian and paired-pulse facilitation results.

Stable negative capacitance effect has recently emerged as a phenomenon that can enable sub 60mV/dec gain in FeFET. Most of the literature reporting on genuine negative capacitance effects make use of epitaxial perovskite films. [125] This finds its origin on the strong dependence of stable negative capacitance, and the nucleation and propagation of ferroelectric domains, which is intrinsically linked to the film quality. However, epitaxial $HfO_2$ films have not been explored on this regard so far. Up to now, it has been shown that strain imposed by the top electrode in a polycrystalline doped hafnium oxide film can have a relevant role on the fine tuning of negative capacitance effects[126] being possible to achieve sub-60mV/dec gain with extremely small hysteresis. Being epitaxial strain highly controllable by the use of f.i. piezoelectric substrate, it is of the highest interest the study of negative capacitance effects in epitaxial films. On the other hand, spatial homogeneity is required, as recently emphasized by Hoffmann et al.[127] Epitaxial HZO films grown on selected substrates ($GdScO_3$ or $TbScO_3$), present pure orthorhombic phase.[64] The recent demonstration of ferroelectricity in ultrathin ferroelectric (epitaxial HZO) / paraelectric (STO or $AlO_x$) on $GdScO_3$ substrates[128] suggests that $HfO_2$ epitaxial ferroelectric films may be excellent candidates for prototyping and exploring negative capacitance devices.

## 9. Conclusions and outlook

Epitaxial films have been much less investigated than polycrystalline films, but they have already demonstrate that can allow excellent ferroelectric properties, and can help to better understand the properties of ferroelectric $HfO_2$ and, in some cases, they can be a guide to improve properties of polycrystalline films. However, the research on epitaxial films is still incipient and many important questions are still open.

The kinetics effects are critical in the crystallization of the metastable orthorhombic phase in polycrystalline films. The balance between thermodynamics and kinetics in the epitaxial growth of ferroelectric $HfO_2$ films could influence the relative amount of phases or the grain size of the epitaxial orthorhombic phase. Epitaxial films are grown primarily by PLD. This technique allows the thermodynamics / kinetics balance to be controlled by adjusting the growth temperature, growth rate (instantaneous or average), and plasma energy (by laser fluence, target substrate distance or gas processing pressure). These studies have yet to be addressed. In this regard, it is also important obtaining epitaxial films by other deposition techniques, including MBE and ALD. Relevant results achieved with epitaxial films include the growth of Y-doped $HfO_2$ with controlled out-of-plane orientation and the estimation of their polarization and Curie temperature. It has been also reported that epitaxial Y-doped $HfO_2$ films as thick as around 1 µm can be ferroelectric, without the degradation of polarization usual in films thicker than a few tens of nanometers. Epitaxial $Hf_{0.5}Zr_{0.5}O_2$ films have evidenced a very weak wake-up effect, contrary to polycrystalline that generally exhibit pronounced wake-up. Epitaxial $Hf_{0.5}Zr_{0.5}O_2$ films have also allowed achievement of simultaneous high polarization, retention, and endurance up to $10^{11}$ cycles, demonstrating that there is not intrinsic dilemmas between P and endurance and between endurance and retention. In contrast, epitaxial $Hf_{0.5}Zr_{0.5}O_2$ appears to suffer more pronounced fatigue. It could be favored because the coercive field of epitaxial films is higher than in polycrystalline films. Understanding the cause of the higher coercive field, and identifying the mechanisms of fatigue in epitaxial films, requires more research. The control of coercive field could allow improvement of endurance.

Most of the research on epitaxial hafnia has been focused on two specific chemical compositions, $Y_{0.07}Hf_{0.93}O_2$ and $Hf_{0.5}Zr_{0.5}O_2$, using mostly ITO and LSMO bottom electrodes, respectively, and usually top Pt electrodes. The properties of other compositions (dopant atom and content), and the effect of other top electrodes, has to be investigated.

Detailed studies of endurance and retention have been reported for $Hf_{0.5}Zr_{0.5}O_2$ films, but scarcely for other compositions. The influence of the relative amount of the parasitic paraelectric phase on the endurance and retention of epitaxial doped $HfO_2$ films is also unknown.

The ferroelectric phase in polycrystalline films is considered to be orthorhombic $Pca2_1$. In the case of epitaxial films there is not consensus and, while some authors consider that the ferroelectric phase is the orthorhombic $Pca2_1$ too, others have proposed rhombohedral phases (R3m and R3) or the orthorhombic $Pnm2_1$. However, there is an important mismatch between DFT calculations and films properties (required strain for the measured polarization), and there is not evidence of differences in specific intrinsic ferroelectric properties in the epitaxial films respect to polycrystalline ($Pca2_1$) films. Deeper characterization of the epitaxial films would be relevant. For example, imaging of dipoles by scanning transmission electron microscopy. Also, there are not reports on ferroelectric switching dynamics in epitaxial films.

Epitaxial $Hf_{0.5}Zr_{0.5}O_2$ have allowed detangling and controlling electronic and ionic contributions to electroresistance in FeTJs or resistive RAMs. However, epitaxial growth impact on other devices has not been investigated in detail. Regarding, FeTJ or ReRAM although complementary ferroelectric and electroresistance characterization signals their connection, resistance switching at faster time scales and their retention and endurance characterization are needed to set potential applications of these type of devices. Fabrication of Fe-FETs based in epitaxial $HfO_2$, and the evaluation of negative capacitance effects in these devices, has not done yet. Near-pure orthorhombic epitaxial HZO films deposited on LSMO/$GdScO_3$ or LSMO/$TbScO_3$ may be useful for prototyping Fe-FETs and exploring negative capacitance effects.

The progress achieved in few years on epitaxial $HfO_2$ films has been outstanding. It is however in a nascent state and probably epitaxial films will be very important in the development of the scientifically exciting and technologically highly relevant ferroelectric $HfO_2$.

**Acknowledgements**

Financial support from the Spanish Ministry of Science and Innovation, through the Severo Ochoa FUNFUTURE (CEX2019-000917-S), MAT2017-85232-R (AEI/FEDER, EU), PID2020-112548RB-I00 (AEI/FEDER, EU), and PID2019-107727RB-I00 (AEI/FEDER, EU) projects, from CSIC through the i-LINK (LINKA20338) program, and from Generalitat de Catalunya (2017 SGR 1377) is acknowledged. Project supported by a 2020 Leonardo Grant for Researchers and Cultural Creators, BBVA Foundation. IF acknowledges Ramón y Cajal contract RYC-2017-22531.

## Figures

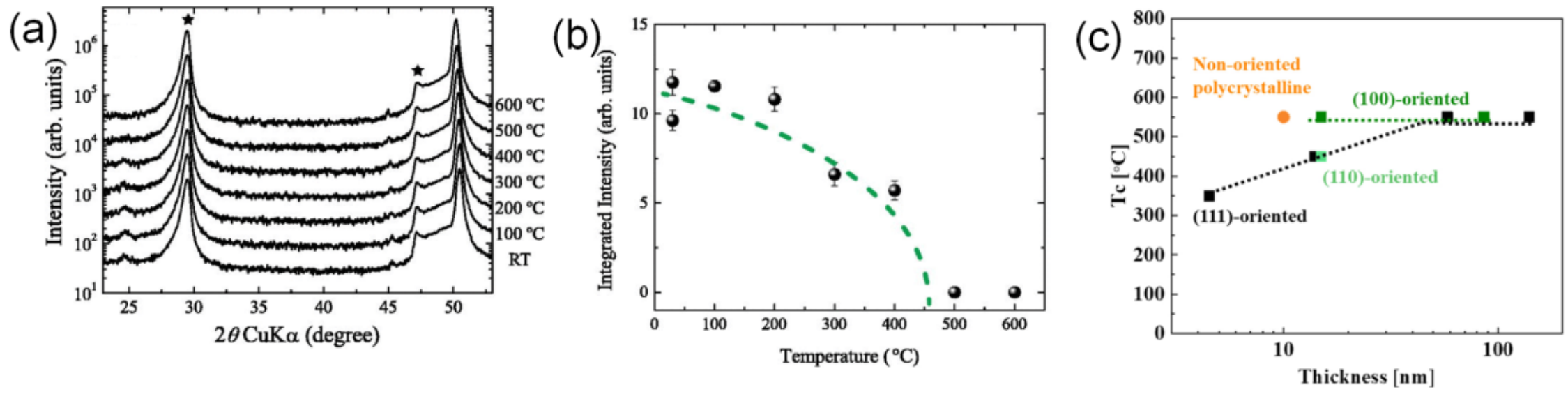

*Figure 1. (a) The XRD patterns with inclination angle of 45° observed for $Hf_{0.93}Y_{0.07}O_2$ film measured from room temperature to 600 °C. (b) The integrated intensity of the 110 super-spot of $Hf_{0.93}Y_{0.07}O_2$ film as a function of temperature. Adapted from ref. 33. Copyright 2015 AIP Publishing. (c) Phase transition temperature, $T_C$, as a function film thickness for oriented and randomly-oriented $Hf_{0.93}Y_{0.07}O_2$ films. Adapted from ref. 39. Copyright 2020 Japan Society of Applied Physics.*

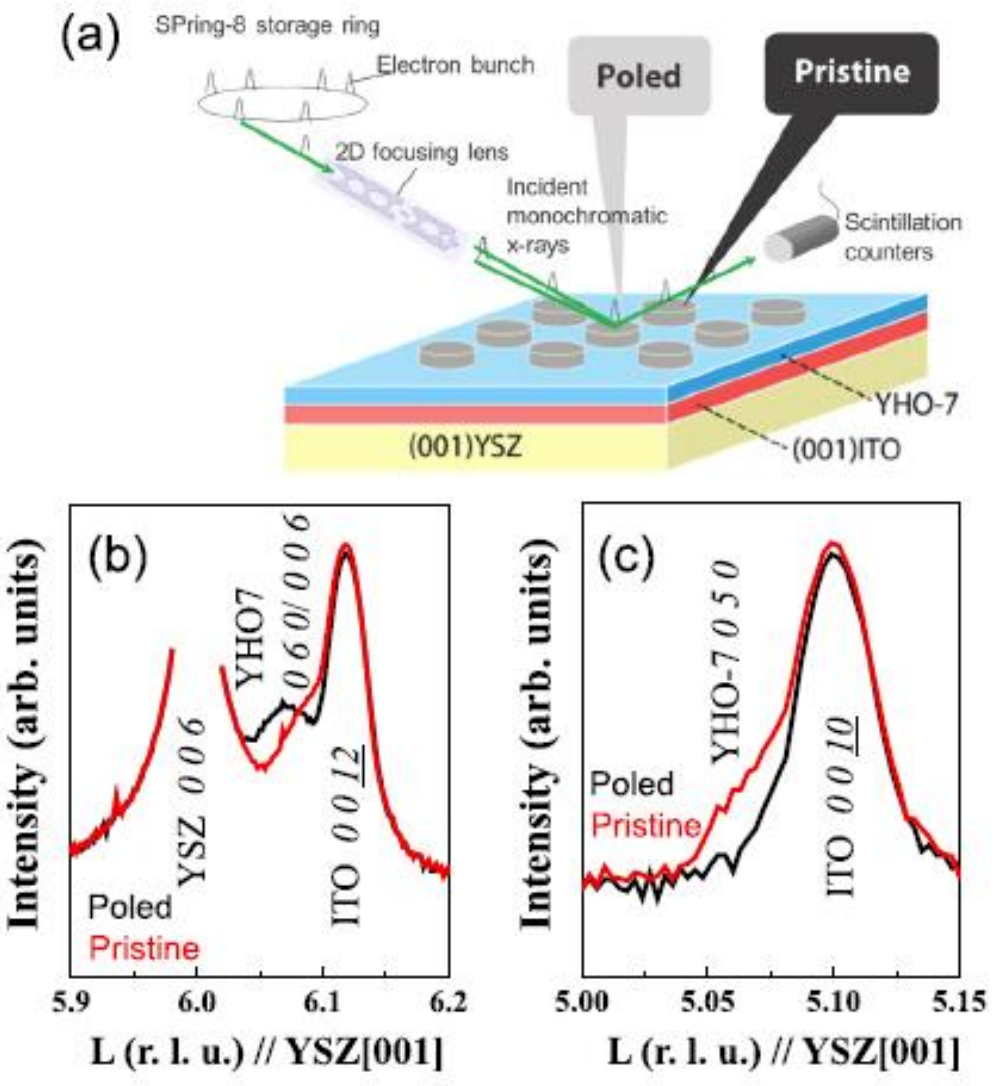

*Figure 2. (a) Schematic diagram for the synchrotron micro-beam X-ray diffraction measurements. XRD pattern scanned along the L-axis in reciprocal space in the vicinity of (b) L = 6 and (c) L = 5. L-axis is parallel to the YSZ [001] direction. Adapted from ref. 36. Copyright 2018 AIP Publishing.*

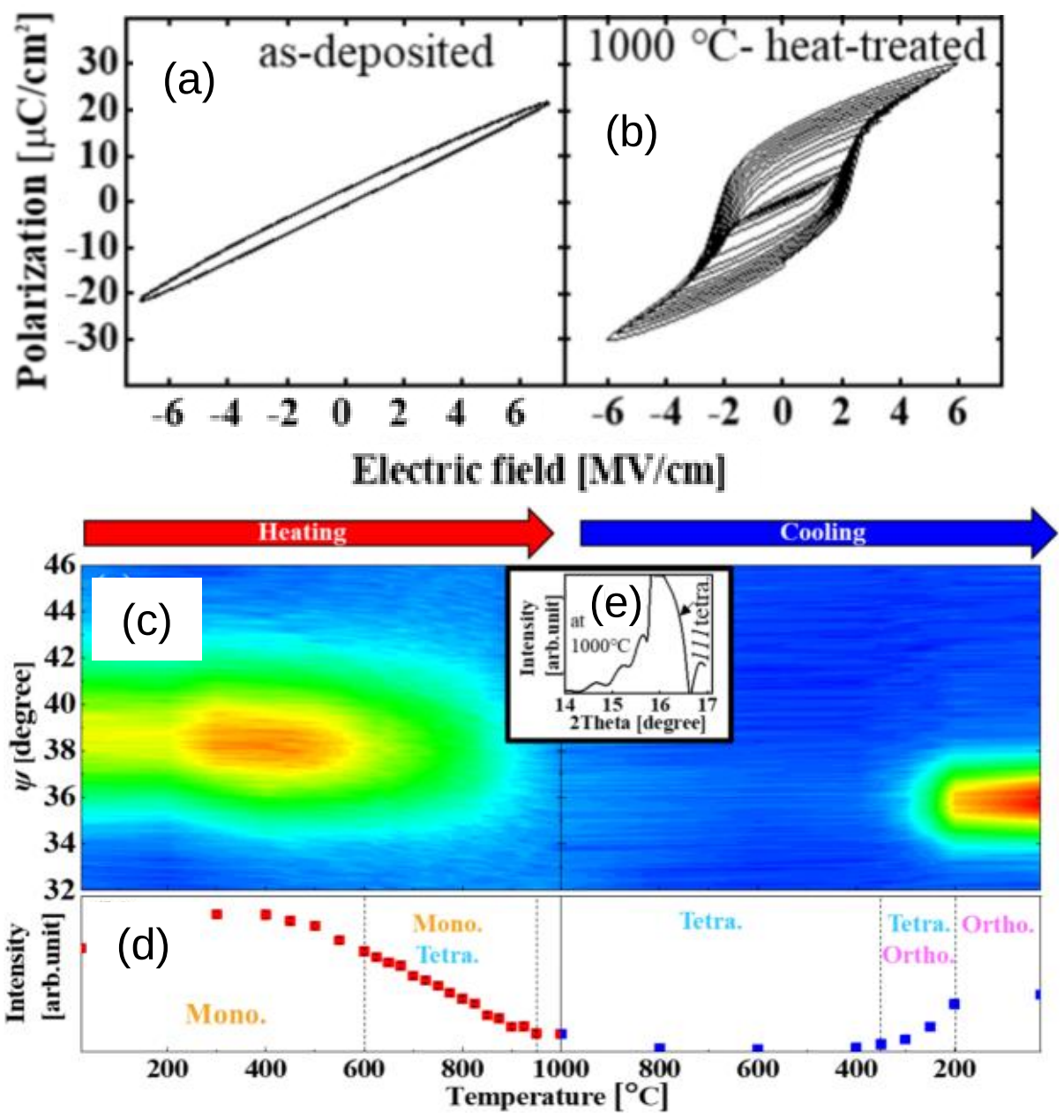

*Figure 3. P–E hysteresis loops measured at 10 kHz for (a) as-deposited and (b) heat-treated $Hf_{0.93}Y_{0.07}O_2$ films on (111)ITO/(111)YSZ substrate; (c) In situ high-temperature XRD result for as-deposited $Hf_{0.93}Y_{0.07}O_2$ film measured at various temperatures in the range from 30 °C to 1000 °C by ψ scanning at $2\theta = 15.7°$ (λ = 0.0827 nm); (d) change in peak intensity during heating and cooling from in-situ high-temperature XRD; (e) XRD $\theta$–$2\theta$ profiles for $Hf_{0.93}Y_{0.07}O_2$ film at 1000 °C (λ = 0.0827 nm). Adapted from ref. 44. Copyright 2019 Japan Society of Applied Physics.*

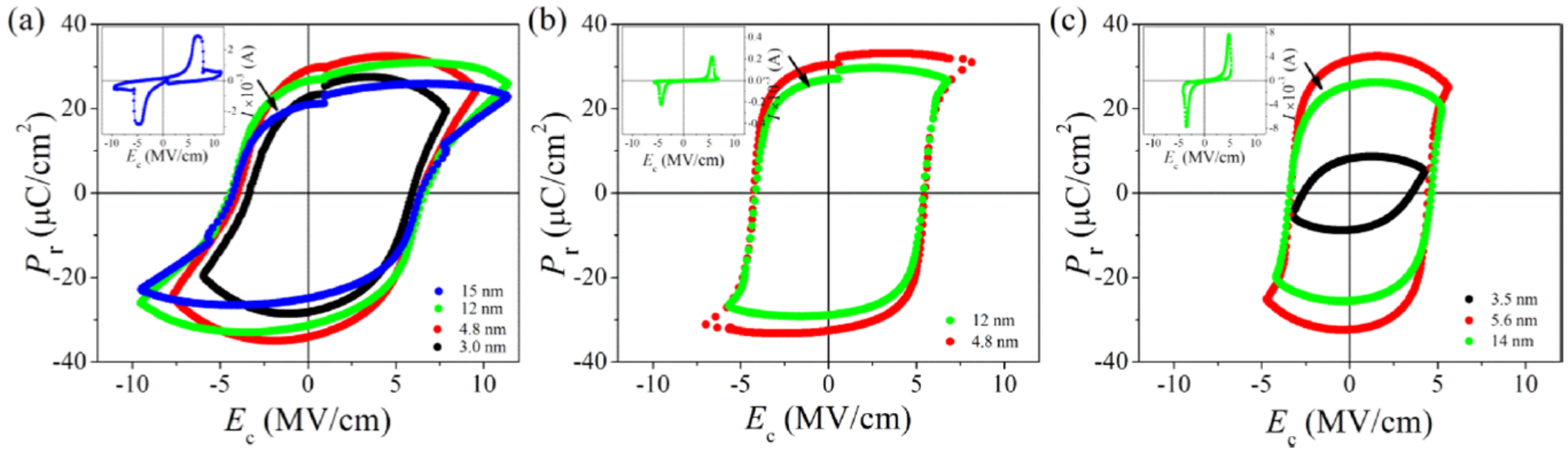

*Figure 4. Ferroelectric polarization loops of Si-doped $HfO_2$ films of different thickness, indicated at bottom right of each panel, grown on (a,b) Nb:STO(110) and (c) Nb:STO(111) and with top Cr/Au electrodes of different area. Adapted from 50. Copyright [2019] American Chemical Society*

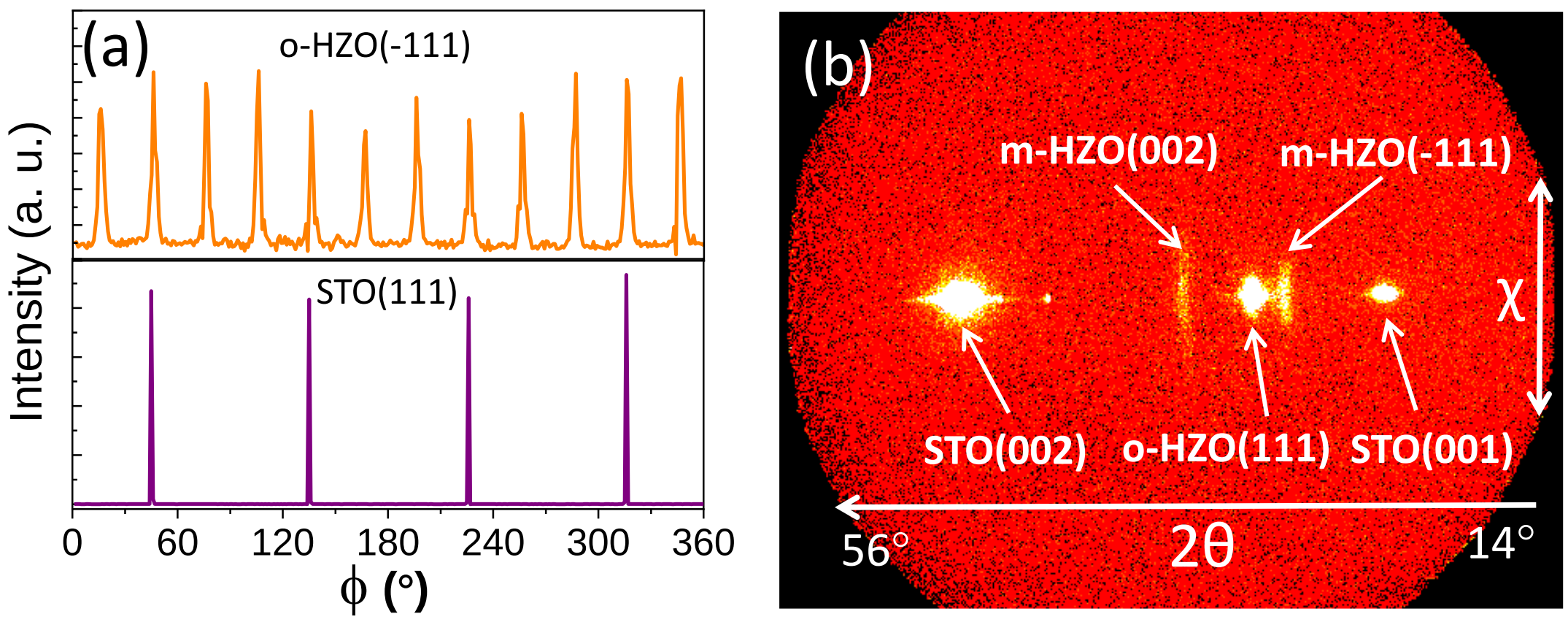

*Figure 5. (a) XRD $\phi$-scans around o-HZO(-111) (top panel) and STO(111) (bottom panel) asymmetrical reflections and (b) $2\theta$-$\chi$ frame of a t = 18 nm HZO film on LSMO/STO(001).. Adapted from ref. 52.Copyright 2018 AIP Publishing*

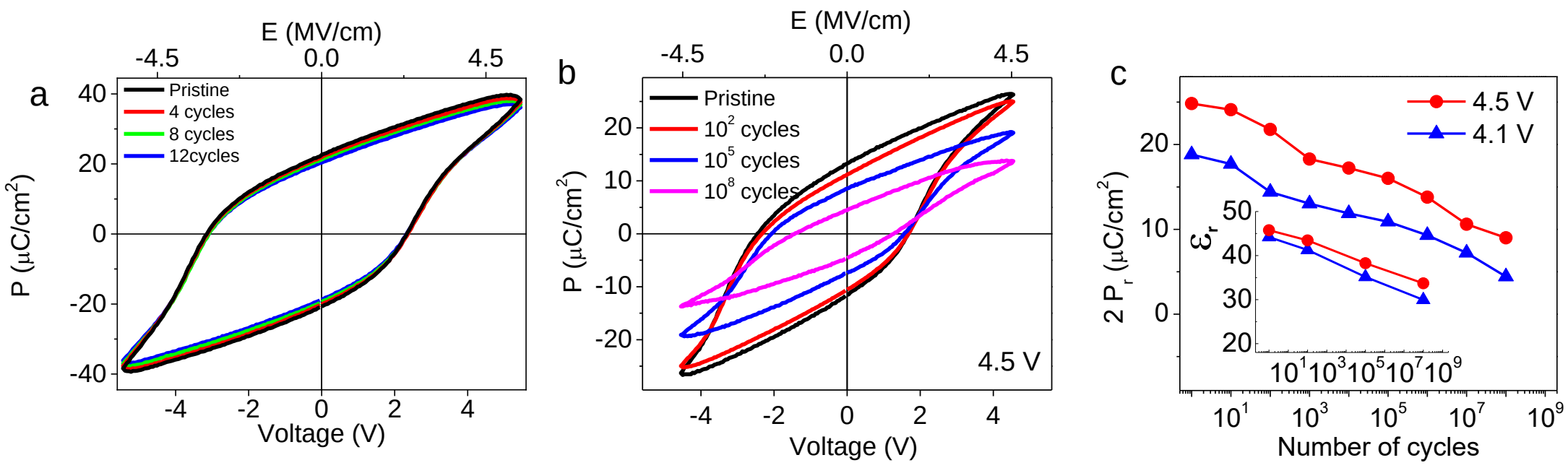

*Figure 6. (a) Ferroelectric hysteresis loops (t = 9 nm film) for the pristine state and subsequently measured cycles. (b) Ferroelectric hysteresis loops (t = 9 nm film) recorded at 1 kHz, for the pristine state and after indicated number of electric cycles at 10 kHz at 4.5 V. (c) Remnant polarization versus number of cycles at 4.5 and 4.1 V. Inset: Dielectric permittivity versus number of cycles at 4.5 and 4.1 V. Adapted from ref. 52. Copyright 2018 AIP Publishing.*

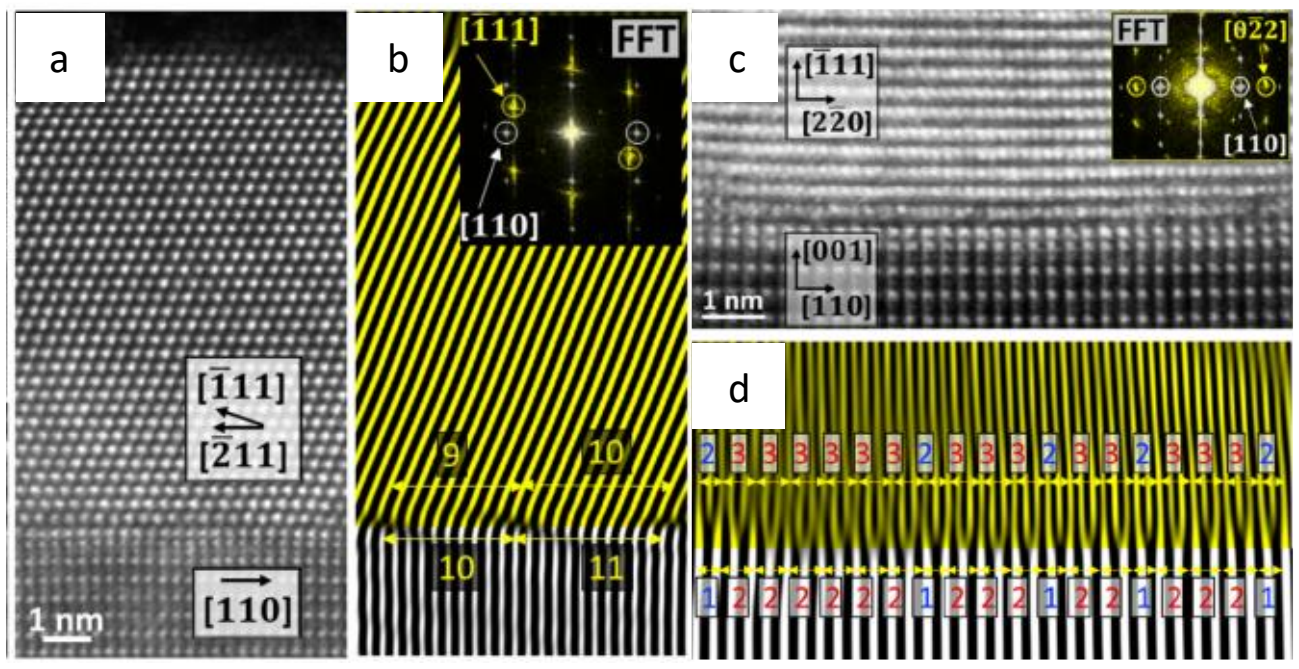

*Figure 7. (a) Cross-sectional STEM image of the HZO/LSMO heterostructure showing a HZO[-211]/LSMO[1-10] crystal variant. (b) Reconstructed image from reflections in the Fourier space corresponding to [−111] HZO and [110] LSMO planes. (inset) The fast-Fourier transform (FFT) of both HZO and LSMO. For the sake of clarity, planes in the HZO layer are shown in yellow, while planes in the LSMO are white. Two adjacent 9/10 and 10/11 domains are indicated. (c) Cross-sectional STEM image showing a HZO[-211]/LSMO[12-20] crystal variant. (d) Equivalent filtered image extracted from (c) by only considering the FFT marked reflections in the inset in (c). 3/2 (red) and 2/1 (blue) domains are visible. Adapted from ref. 57. Copyright 2020 American Chemical Society.*

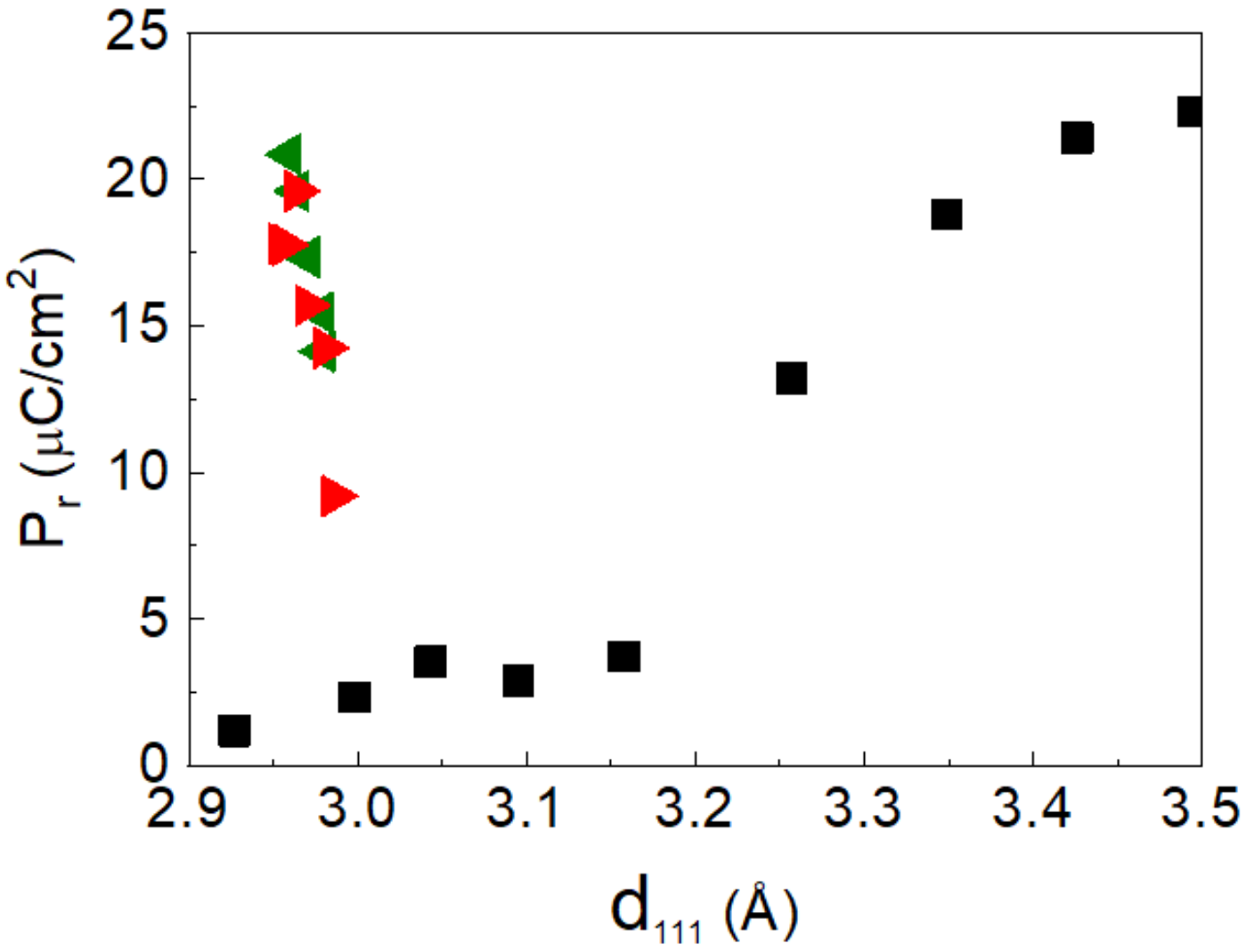

*Figure 8. Experimental data of remanent polarization and out-of-plane lattice parameter of epitaxial HZO thin films, around 9 nm thick, deposited on LSMO/STO(001) under varying temperature (green triangles) and oxygen pressure (red triangles). Figure adapted from ref. 53. Copyright 2019 American Chemical Society. Black squares: computed polarization of the R3m phase of HZO as a function of $d_{111}$ (Figure adapted from ref. 51. Copyright Springer Nature 2018)).*

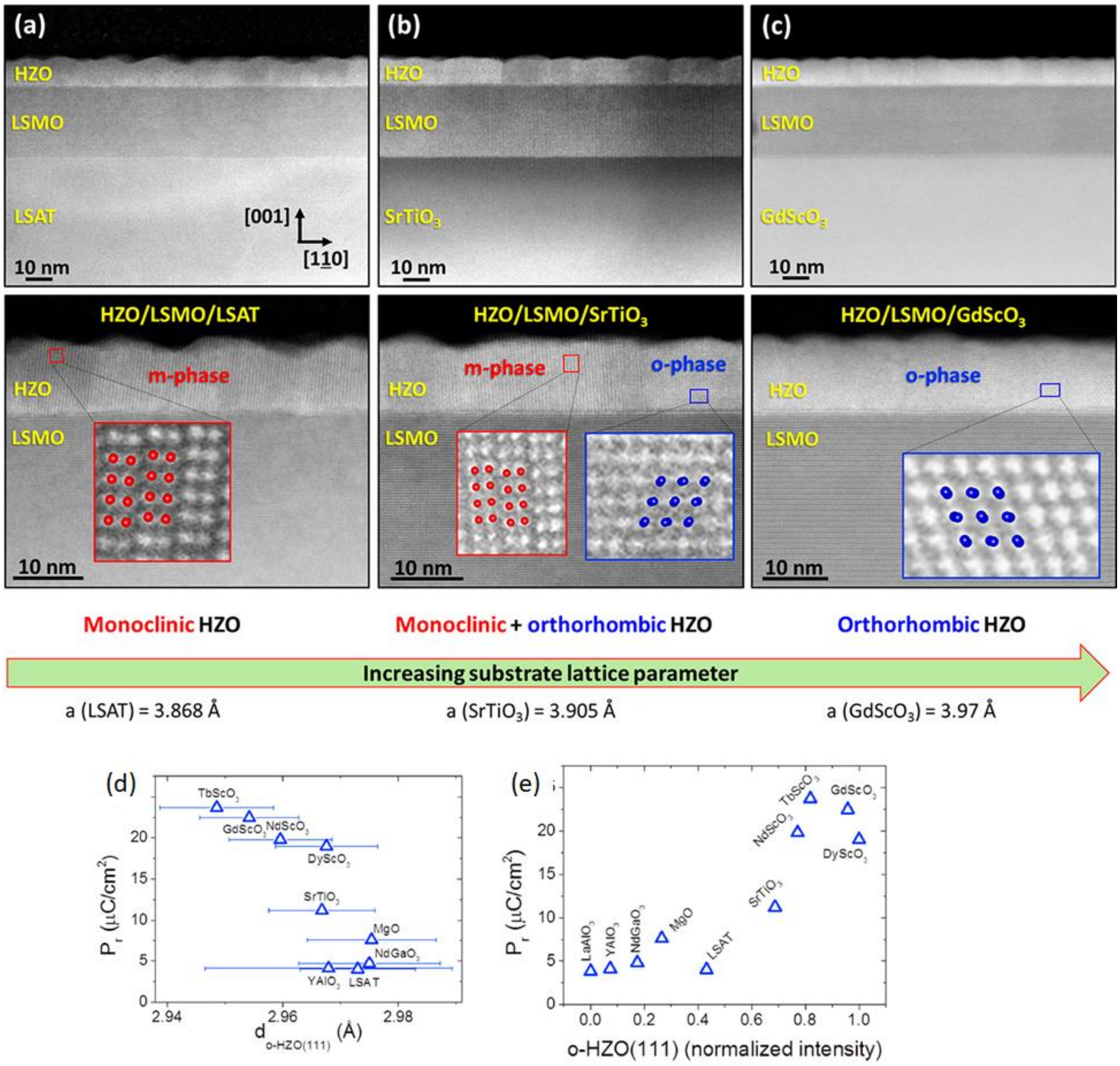

*Figure 9. Cross-sectional STEM images of HZO/LSMO films on (a) LSAT, (b) $SrTiO_3$, and (c) $GdScO_3$ substrates. Top panels are low-magnification images showing the substrate, the LSMO, and the HZO films. Bottom panels are higher magnification images of the HZO and LSMO films. The insets show atomic-resolution images of the HZO films. Red and blue circles depict the monoclinic (space group $P2_1/c$) and orthorhombic (space group $Pca2_1$) structures, respectively. Remanent polarization as a function of the interplanar $d_{o\text{-}HZO(111)}$ spacing. $d_{o\text{-}HZO(111)}$ was determined by Gaussian fits of the XRD $2\vartheta$ peak position, and the error bar is set to $1\sigma$ of the fit. Remanent polarization as a function of the o-(111) out-of-plane lattice distance (d) and as a function of the normalized intensity of the XRD o-HZO(111) reflection (e). Adapted from ref. 64. Copyright 2019 American Chemical Society.*

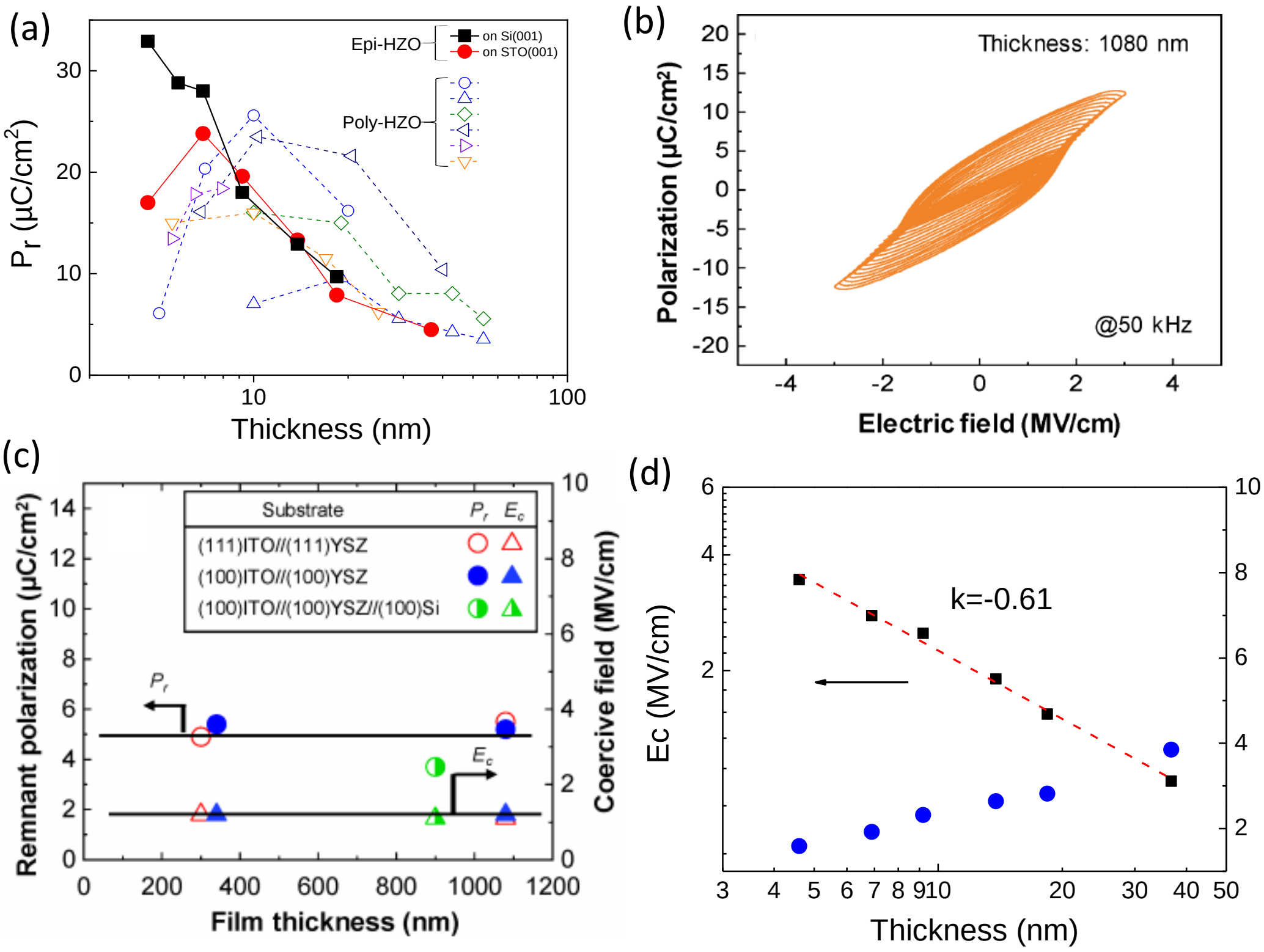

*Figure 10. (a) Remanent polarization (black solid squares) plotted as a function of the thickness of epitaxial HZO films on LSMO/LaNiO3/$CeO_2$/YSZ/Si(001). Red solid circles show the polarization of epitaxial films on LSMO/$SrTiO_3$(001). Empty symbols correspond to several thickness series reported in literature for polycrystalline HZO films. Adapted from ref. 85. Copyright 2020 Royal Society of Chemistry. References corresponding to plotted data can be found there. (b) Room-temperature P-E hysteresis loops measured at 50 kHz for a t = 1080 nm $Hf_{0.93}Y_{0.07}O_2$ film on ITO(001)/YSZ(001) (c) Film-thickness dependence of remanent polarization (circles) and coercive field (triangles) for $Hf_{0.93}Y_{0.07}O_2$ films on ITO(111)/YSZ(111) (open symbols), ITO(001)/YSZ(001)YSZ (closed symbols) and ITO(001)/YSZ(001)/Si(001) (half-closed symbols) substrates. (b,c) Adapted from ref. [91]. Copyright 2020 Japan Society of Applied Physics. (d) Dependences of $E_C$ (black squares) and $V_C$ (blue circles) on thickness of epitaxial HZO films on LSMO/$SrTiO_3$(001). The red dashed line is a linear fit with a slope of −0.61, compatible with $E_C – t^{-2/3}$ scaling. Adapted from ref. 53. Copyright 2019 American Chemical Society.*

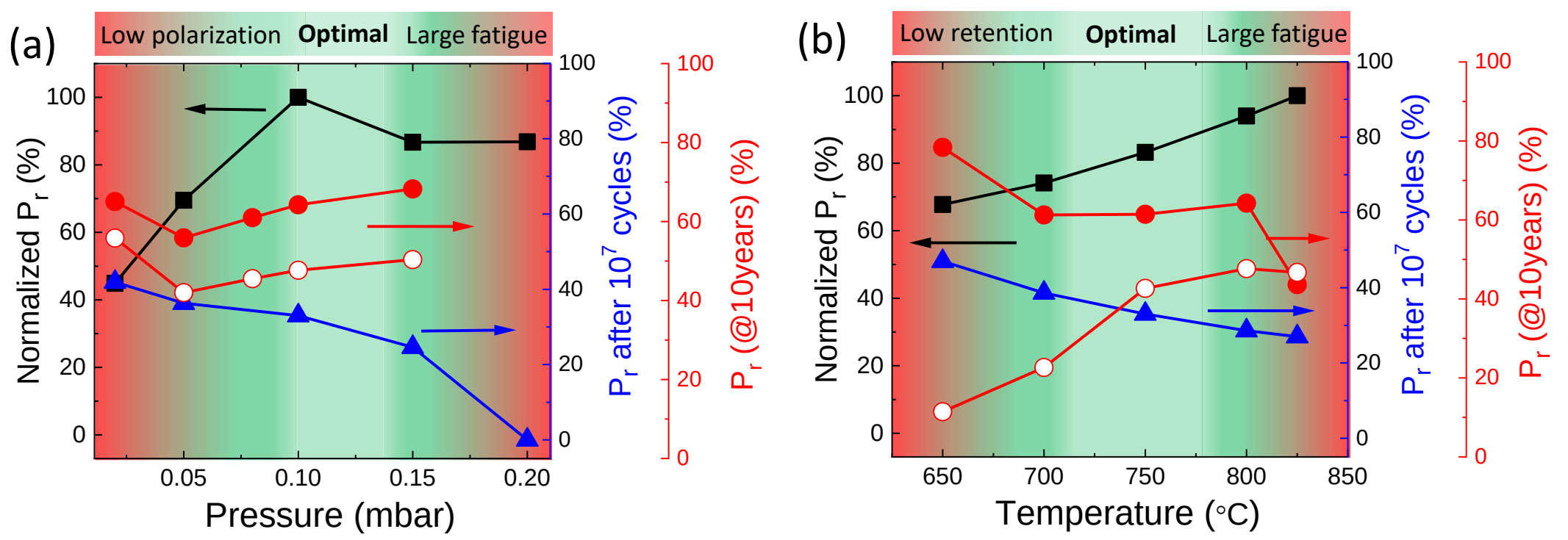

*Figure 11. Dependences of the normalized remanent polarization in the pristine state (black squares, left axis) and remanent polarization after $10^7$ cycles (blue triangles, right axis), and normalized remanent polarization after 10 years of positive (red solid circles, right axis) and negative (red empty circles, right axis) poling as a function of (a) oxygen pressure and (b) substrate temperature. Colored areas summarize the main effect of the deposition parameters on polarization, fatigue, and endurance. Adapted from ref. 84. Copyright 2020 AIP Publishing.*

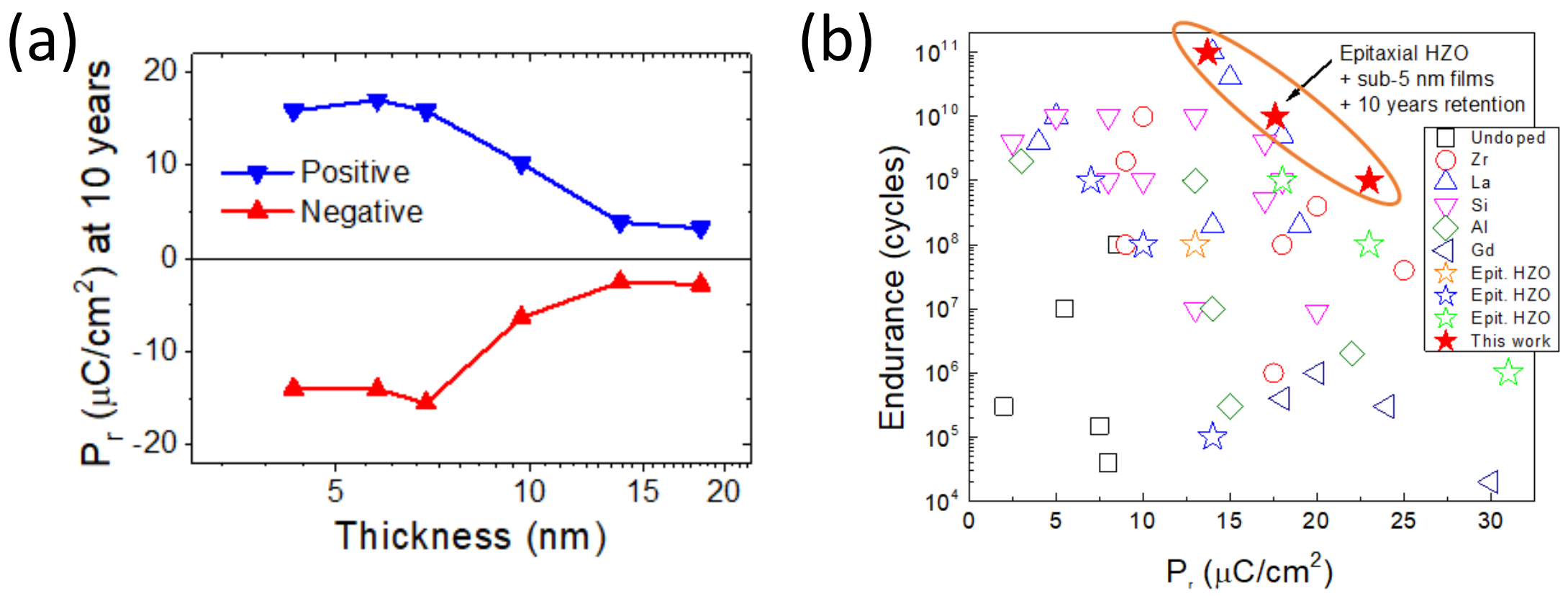

*Figure 12. (a) Extrapolated remnant polarization at 10 years using the data fitted to the Pr = P0td−k equation in HZO/LSMO/LaNiO3/CeO2/YSZ/Si(001) samples. (b) Endurance and remnant polarization reported in literature for HfO2 films undoped or with various dopants. The remnant polarization is the maximum value (after wake up in the case of polycrystalline films). Stars correspond to epitaxial HZO films. Solid red stars correspond to the t = 4.6 nm epitaxial HZO/LSMO/LaNiO3/CeO2/YSZ/Si(001) samples film, which extrapolated retention is shown in panel (a) . Other symbols correspond to various undoped or doped HfO2 polycrystalline films. Adapted from ref. 85. Copyright 2020 Royal Society of Chemistry. References corresponding to plotted data can be found there.*

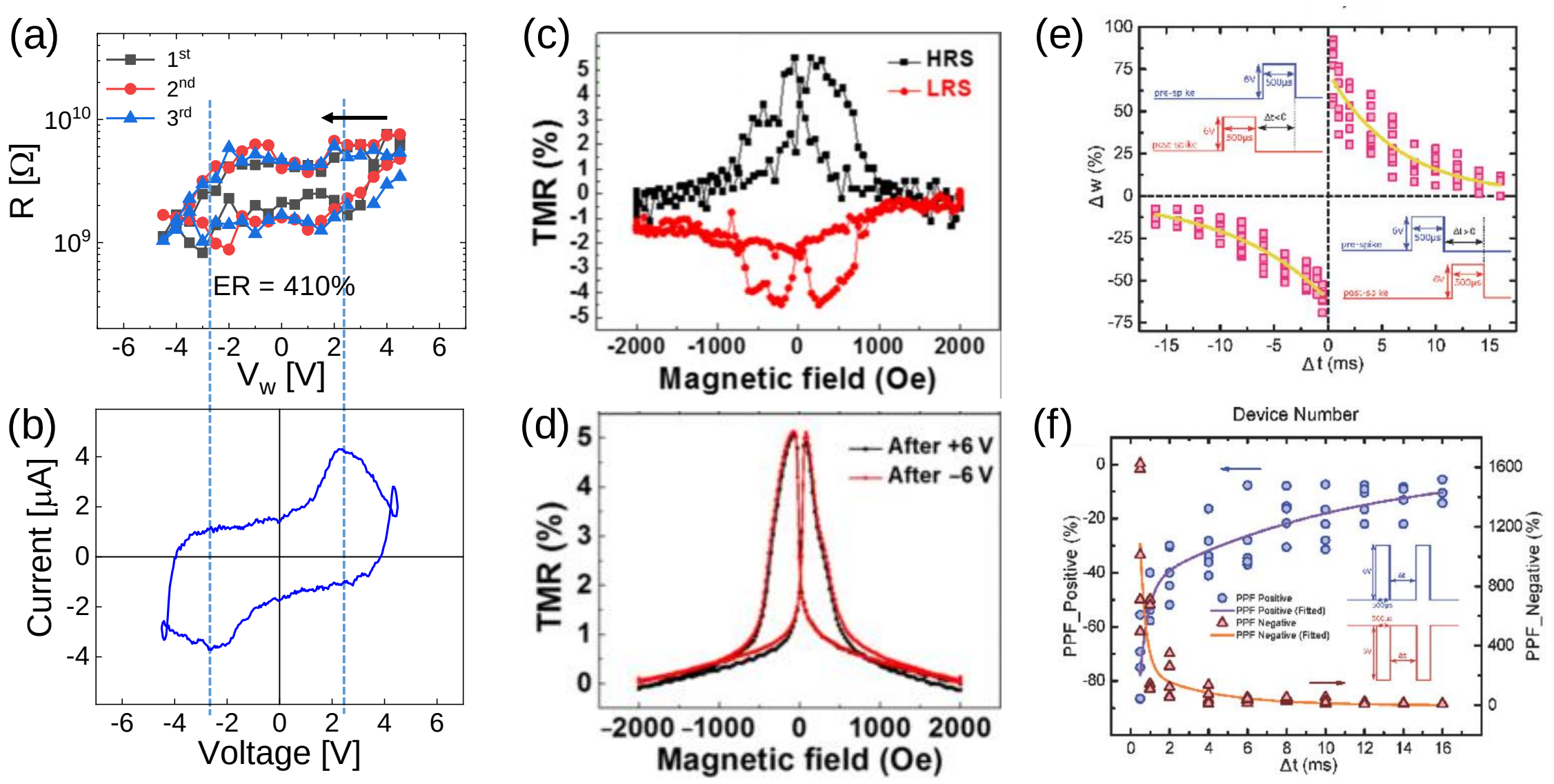

*Figure 13. (a) Dependence of the junction resistance (R) on the writing voltage (VW) on Pt/HZO(t=4.6 nm)/LSMO capacitor on STO(001). (b) I–V loop of same sample measured just before performing R(VW) measurement cycles. (a-b) are adapted from ref. 76. Copyright 2020 John Wiley & Sons, Inc. (c) TMR ratio as a function of sweeping magnetic field in the high resistance state (HRS, black squares) and low resistance state (LRS, red circles) measured after repeated application of ±6 V electric pulses. Measurements are performed at 50 K on a junction device with an electrode area of 30 × 30 µm2 of a Co/HZO/LSMO/STO(001) sample. Adapted from ref. 83. (d) TMR as a function of the magnetic field under a bias of −0.2V at 50 K. Adapted from ref. 81. Copyright 2019 American Physical Society (e) Asymmetric Hebbian learning rule and fitted results in solid yellow line. (f) Measured paired-pulse facilitation (PPF) ratios of positive and negative pulse trains and fitted results in purple and orange solid lines, respectively. The pulse waveforms for all measurements are included in the insets for (e) and (f). (e-f) are adapted from ref. 78. Copyright 2018 John Wiley & Sons, Inc.*

## Tables

*Table I: summary of main ferroelectric properties reported for epitaxial films. * symbol means that the corresponding value was not indicated. value. Scandate substrates are indexed as pseudocubic. A similar Table, summarizing ferroelectric properties for polycrystalline doped $HfO_2$ films, is reported in reference 31.*

| Material | Deposition method | Deposition temperature (°C) | Deposition atmosphere | Substrate | Top / bottom electrode | Thickness (nm) | $P_r$ (μC/cm$^2$) | $E_c$ (MV/cm) | Endurance (cycles) / $E_{cycling}$ (MV/cm) | Retention / $E_{poling}$ (MV/cm) | Reference |
|---|---|---|---|---|---|---|---|---|---|---|---|
| $Hf_{0.93}Y_{0.07}O_2$ | PLD | 700 °C | 0.01 Torr ($O_2$) | YSZ(110) | Pt / ITO | 15 | ~12 | ~2 | - | - | Shimizu[37] |
| $Hf_{0.93}Y_{0.07}O_2$ | PLD | 700 °C | 0.01 Torr ($O_2$) | YSZ(111) | Pt/ ITO | 14 | ~10 | ~2 | - | - | Katayama[38] |
| $Hf_{0.93}Y_{0.07}O_2$ | Sputtering | RT + 1000 °C annealing | 0.2 Torr (Ar) | YSZ(111) | Pt / ITO | 24 | ~11 | ~2.2 | - | - | Suzuki [45] |
| $Hf_{0.93}Y_{0.07}O_2$ | PLD | RT + 1000 °C annealing | 0.01 Torr ($O_2$) | YSZ(111) | Pt / ITO | 15 | 15 | ~2.1 | - | - | Mimura[44] |
| $Hf_{0.93}Y_{0.07}O_2$ | PLD | RT + 1000 °C annealing | 0.01 Torr ($O_2$) | YSZ(111) | * / ITO | 111 | ~5 | ~1.4 | - | - | Mimura[80] |
| $Hf_{0.93}Y_{0.07}O_2$ | Sputtering | RT + 800 °C annealing | 0.2 Torr (Ar) | YSZ(111) and (001) | * / ITO | 380 and 1080 | ~5 | ~1 | - | - | Shimura[91] |
| $Hf_{0.94}Fe_{0.06}O_2$ | Ion beam | RT + 900 °C annealing | 3.8x10$^{-5}$ Torr | YSZ(001) | Pt / ITO | 20 | 8.8 | ~2 | - | - | Shiraishi[41] |
| $Hf_{0.9}Ce_{0.1}O_2$ | Ion beam | RT + 900 °C annealing | 3.8x10$^{-5}$ Torr | YSZ(001) | Pt / ITO | 30 | ~5 | - | - | - | Shiraishi[42] |
| $Hf_{0.93}Y_{0.07}O_2$ | Sputtering | RT | 0.01 Torr (Ar) | YSZ(111) | Pt / ITO | 16 | 15 | 2.3 | - | - | Mimura[48] |
| $Hf_{1-x}Y_xO_2$ (x=*) | PLD | 700 °C | 0.15 Torr ($O_2$) | YSZ/Si(001) | Pt / - | * | ~20 (leaky) | * | - | - | Lee[90] |
| $Hf_{0.936}Si_{0.044}O_2$ | PLD | 700 °C | 0.1 Torr ($O_2$) | Nb:STO(111) and (110) | Au-Cr / substrate | 3 - 15 | up to ~32 | 4-5 | - | - | Li[50] |
| $Hf_{0.5}Zr_{0.5}O_2$ | PLD | 700 °C | - | YSZ(111) and (110) | Au-Cr / TiN | 15 | ~7-20 | 1.1-2.3 | - | - | Li[40] |
| $Hf_{0.5}Zr_{0.5}O_2$ | PLD | 800 °C | 0.1 mbar | STO(001) | Pt / LSMO | 9 | 20 | 3 | $1x10^8$ (5) | >10 (6.1) | Lyu[52] |
| $Hf_{0.5}Zr_{0.5}O_2$ | PLD | 800 °C | 0.1 mbar | STO(001) | LSMO / LSMO | 5 | 34 | ~5 | - | - | Wei[51] |
| $Hf_{0.5}Zr_{0.5}O_2$ | PLD | 800 °C | 0.1 mbar | STO(001) | LSMO / LSMO | 9 | 18 | ~3 | $1x10^5$ (4.4) | - | Wei[51] |
| $Hf_{0.5}Zr_{0.5}O_2$ | PLD | 550 °C | 0.13 mbar | LAO(001) | Pd / LSMO | 10 | 20 | 2.4 | - | - | Yoong[78] |
| $Hf_{0.5}Zr_{0.5}O_2$ | PLD | 800 °C | 0.1 mbar | YSZ/Si(001) | Pt / LSMO | 4.6 | 33 | ~4 | $1x10^{11}$ (5.4) | >10 (5.4) | Lyu[116] |
| $Hf_{0.5}Zr_{0.5}O_2$ | PLD | 800 °C | 0.1 mbar | STO/Si(001) | Pt / LSMO | 7.7 | 34 | ~3 | $1x10^9$ (5.2) | >10 (5.2) | Lyu[66] |
| $Hf_{0.5}Zr_{0.5}O_2$ | PLD | 800 °C | 0.1 mbar | $GdScO_3$ and $TbScO_3$(001) | Pt / LSMO | 9 | ~24 | ~2.5 | - | - | Estandia[64] |
| $Hf_{0.5}Zr_{0.49}La_{0.01}O_2$ | PLD | 800 °C | 0.1 mbar | STO(001) | Pt / LSMO | 4.8 | ~20 | ~3.7 | $5x10^{10}$ (5.4) | >10 (5.4) | Song[63] |
| $Hf_{0.5}Zr_{0.49}La_{0.01}O_2$ | PLD | 800 °C | 0.1 mbar | STO/Si(001) | Pt / LSMO | 6.3 | ~30 | ~3.5 | $1x10^9$ (4.3) | >10 (7.2) | Song[63] |
| $Hf_{0.945}La_{0.055}O_2$ | PLD | 600 °C | 0.1 mbar | STO(001) | Pt / LSMO | 12 | ~16 | ~2.7 | $2x10^7$ (5.3) | >10 (5.3) | Li[88] |

***Table II**: Comparison of wake-up effect relevance, fatigue, endurance, retention, and coercive electric field values and its thickness dependence in polycrystalline and epitaxial ferroelectric films of doped $HfO_2$.*

| | **Polycrystalline films** | **Epitaxial films** |
|---|---|---|
| **Wake-up** | Important | Negligible or very low |
| **Fatigue** | Can ve low or null | Severe |
| **Endurance** (best reported) | $10^{11}$ cycles | $10^{11}$ cycles |
| **Retention** (best reported) | >10 years | >10 years |
| **$E_c$** (usual values) | ~ 1 – 2 MV/cm | ~ 2 – 3 MV/cm |
| **$E_c$ versus thickness** | Little dependence | $E_c$ – $t^{-2/3}$ scaling |